\documentclass[11pt]{article}

\usepackage{graphicx}
\usepackage{color}
\usepackage{latexsym}
\usepackage{amsmath,amsfonts,amssymb}

\newcommand{\be}{\begin{equation}}
\newcommand{\ba}{\begin{eqnarray}}
\newcommand{\ee}{\end{equation}}
\newcommand{\ea}{\end{eqnarray}}

\newcommand{\barr}{\begin{array}}
\newcommand{\ear}{\end{array}}
\newcommand{\ns}{\normalsize}

\newcommand{\di}{d}        

\newcommand{\ax}{\alpha}
\newcommand{\bx}{\beta}
\newcommand{\cx}{\gamma}
\newcommand{\dx}{\delta}
\newcommand{\ex}{\epsilon}
\newcommand{\ox}{\omega}

\newcommand{\ab}{\bar\alpha}
\newcommand{\bb}{\bar\beta}
\newcommand{\cb}{\bar\gamma}

\newcommand{\Ox}{\Omega}

\newcommand{\te}{\tilde \epsilon}
\newcommand{\tm}{\tilde \mu}

\newcommand{\wg}{\wedge}
\renewcommand{\Re}{\mathrm{Re}}
\renewcommand{\Im}{\mathrm{Im}}

\newcommand{\nn}{\nonumber}
\newcommand{\tox}{\tilde\omega}

\newcommand{\cK}{\mathcal{K}}
\newcommand{\cM}{\mathcal{M}}

\newcommand{\cG}{\mathcal{G}}
\newcommand{\cH}{\mathcal{H}}
\newcommand{\cQ}{\mathcal{Q}}
\newcommand{\Oh}{\mathcal{O}}

\newcommand{\cX}{\mathcal{X}}
\newcommand{\cW}{\mathcal{W}}
\newcommand{\cko}{e^{-K^{(Z)}}} 

\numberwithin{equation}{section}

\begin{document}

\begin{titlepage}

\title{
  \hfill{\normalsize YITP 05-35\\}
  \hfill{\normalsize CERN-TH-PH/2005-128\\[-2mm]} 
  \hfill{\normalsize hep-th/0507173}
  \vskip 2cm
   {\Large\bf Moduli Stabilisation in Heterotic String Compactifications}\\[0.5cm]}
   \setcounter{footnote}{0}
\author{{\ns\large
    Beatriz de Carlos$^1$\footnote{email:
   B.de-Carlos@sussex.ac.uk} $^{,}$\footnote{Also at Department of
   Physics CERN, Theory Division, 1211 Geneva 23, Switzerland.}}~,
{\ns\large
    Sebastien Gurrieri$^2$\footnote{email: gurrieri@yukawa.kyoto-u.ac.jp}}~,
\setcounter{footnote}{3}
{\ns\large Andr\'e Lukas$^3$\footnote{email: lukas@physics.ox.ac.uk}}~
  {\ns and Andrei Micu$^1$\footnote{email: A.Micu@sussex.ac.uk}
   $^{\, ,}$\footnote{On leave from IFIN-HH Bucharest.}}
  \\[0.5cm]
   {\it\ns $^1$Department of Physics and Astronomy, University of Sussex}\\
   {\ns Brighton BN1 9QH, UK}\\[0.5cm]
   {\it\ns $^2$Yukawa Institute for Theoretical Physics, Kyoto University}\\
   {\ns Kyoto 606-8502, Japan} \\[0.5cm]
    {\it\ns $^3$Rudolf Peierls Centre for Theoretical Physics, University of Oxford}\\
   {\ns 1 Keble Road, Oxford OX1 3NP, UK}}
\date{}

\maketitle

\begin{abstract}
  In this paper we analyze the structure of supersymmetric vacua
  in compactifications of the heterotic string on certain manifolds
  with SU(3) structure. We first study the effective theories
  obtained from compactifications on half-flat manifolds and show that
  solutions which stabilise the moduli at acceptable values are hard
  to find. We then derive the effective theories associated with
  compactification on \emph{generalised half-flat manifolds}. It is
  shown that these effective theories are consistent with
  four-dimensional $N=1$ supergravity and that the superpotential can
  be obtained by a Gukov-Vafa-Witten type formula. Within these
  generalised models, we find consistent supersymmetric (AdS) vacua at
  weak gauge coupling, provided we allow for general internal gauge
  bundles. In simple cases we perform a counting of such vacua and
  find that a fraction of about $1/1000$ leads to a gauge coupling
  consistent with gauge unification.
\end{abstract}

\thispagestyle{empty}

\end{titlepage}


\section{Introduction}

There is now a considerable body of work on moduli stabilization,
facilitated by flux of the Neveu Schwarz-Neveu Schwarz (NSNS) and
Ramond-Ramond (RR) anti-symmetric tensor fields,
in the context of type II theories. Specifically, within type IIB it
has been shown~\cite{GKP} that a combination of NSNS and RR
three-form flux can stabilize the dilaton and all complex structure
moduli, while the K\"ahler moduli have to be fixed by other effects
such as non-perturbative contributions~\cite{KKLT} or perhaps
higher-order $\alpha '$ corrections~\cite{BBHL}. The consistency of
these procedures, including the interplay between $\alpha'$ and
non-perturbative corrections, was analysed in
Refs~\cite{BB1,CFNOP,BB2,CQS,LRSS,dA1,dA2} 
Within type IIA theories, on the other hand, both
odd and even degree form field strengths are available, so that flux
potentials for complex structure moduli as well as K\"ahler moduli
will typically be generated \cite{TV,LM} (see also Ref.~\cite{GL2} for
$N=1$ models). One may therefore hope that all moduli can be
stabilised by flux in some such models and specific examples have
indeed been found~\cite{DKPZ,DWGKT,VZ,HP,CFI}, although it appears that
in generic models of this kind some flat directions are still left over.

Traditionally, the heterotic string has been considered the most
attractive string theory, with the presence of (preferably $E_8\times
E_8$) ten-dimensional gauge fields leading to a large number of
supersymmetric compactifications with phenomenologically interesting
properties~\cite{CHSW}.  It has also been known for a long time that
heterotic NSNS three-form flux can stabilize all complex structure
moduli of the theory~\cite{DRSW,RW}. More recently, this subject was
addressed in Refs~\cite{GKLMcA,CKL}. However, in the absence of any
further (RR) antisymmetric tensor fields, the potential for
stabilizing the remaining moduli seems rather limited compared to type
II theories. This apparent problem can be overcome by departing from
Calabi-Yau compactifications and by considering the heterotic string
on general manifolds with $\mbox{SU(3)}$ structure.  Such models were
analysed in Refs~\cite{CCDLMZ,BBDG,BBDP,CCDL,BBDGS,ST,BD} where
general aspects of compactifications on non-K\"ahler manifolds were
studied. Recently, a more general analysis which takes into account
the effects of a gaugino condensate in ten dimensions appeared in
Ref.~\cite{FL}. The generic form of the superpotential was inferred in
\cite{BBDG,BBDP,CCDL}, but its detailed analysis was not possible due
to the lack of knowledge of the moduli space of these manifolds.  On
the other hand the low energy effective action and an explicit form
for the superpotential in terms of the low energy fields were found in
Ref.~\cite{GLM}, where half-flat mirror manifolds were used as
compactification spaces. Such manifolds arise in the context of type
II mirror symmetry with NS fluxes \cite{GLMW} and, in some appropriate
region, their moduli space was conjectured to be similar to that of a
normal Calabi--Yau manifold. This conjecture has been applied in
Ref.~\cite{GLM} to derive the low-energy theory for these
compactifications and, in particular, the superpotential as an
explicit function of the moduli fields. In particular, it was found that the
intrinsic torsion of the half-flat mirror manifolds gives rise to a
superpotential for the K\"ahler moduli. These results suggest that, by
combining the intrinsic torsion of sufficiently general classes of
$\mbox{SU}(3)$ structure manifolds with NSNS flux, moduli
stabilization in heterotic compactifications can be as flexible as in
type II models.

This is precisely the line of work we would like to further develop in
the present paper. We will first study in detail supersymmetric moduli
stabilization in the heterotic string on half-flat mirror manifolds,
based on the effective theories of Ref.~\cite{GLM}. As we will see,
the torsion of half-flat manifolds and the allowed $H$-fluxes are
insufficient to fix all K\"ahler and complex structure moduli. We
therefore move on to the more general class of manifolds with
$\mbox{SU}(3)$ structure described in Refs~\cite{DAFTV,AT,GLW}, which
we will refer to as \emph{generalised half-flat manifolds}. For this
class of spaces we first show, by an explicit reduction of the bosonic
action, that the heterotic Gukov-Vafa-Witten type formula for
heterotic compactifications~\cite{GLM,GVW,SG} leads to the correct
result for the superpotential. This result is then applied to a
detailed analysis of moduli stabilization for those models.

The flux and torsion superpotential, ${\cal W}$, for all models
considered in this paper is a function of the K\"ahler moduli $T^i$
and complex structure moduli $Z^a$, but it turns out to be independent
of the dilaton $S$.  Hence, the dilaton is not stabilised at this
stage. However, in the $E_8\times E_8$ case, one expects hidden sector
gaugino condensation to generate a non-perturbative dilaton
superpotential which should be added to ${\cal W}$. We will,
therefore, use this non-perturbative contribution to stabilize the
dilaton. It turns out that, in order to fix $S$ at sufficiently weak
coupling (and to be in the large radius and large complex structure
limits) we need global minima for the K\"ahler and complex structure
moduli which correspond to superpotential values ${\cal W}_0$ with
$|{\cal W}_0| \ll 1$. This is quite analogous to a similar
requirement in type IIB models~\cite{KKLT}, where it is necessary to ensure
moduli stabilization at large radius. The original models of heterotic
gaugino condensation with flux~\cite{DRSW,RW} were discarded precisely
because this condition was difficult to satisfy due to the quantization
of fluxes. However, we find that cancellations leading to small $|{\cal W}_0|$
are possible for our generalised models. We carry out a
{\em statistical} analysis in those cases, counting the number of vacua
as a function of $|{\cal W}_0|$ and the maximal flux value. As the
value of $|{\cal W}_0|$ determines the value of the dilaton, this
counting analysis is directly relevant to the question of how many
vacua realize a phenomenologically acceptable gauge coupling.

The outline of the paper is as follows. In Section~\ref{extension} we
briefly review the low energy effective theory of the heterotic string
on half-flat mirror manifolds~\cite{GLM}. In addition, we work out the
generalization of this effective theory expected for the more general
spaces proposed in~\cite{DAFTV}. We will show explicitly that the
potential obtained from compactification (which includes a part from
the non-vanishing scalar curvature of the internal space) can be
obtained from a Gukov-Vafa-Witten type superpotential for manifolds
with SU(3) structure, which was derived in~\cite{GLM}. In
Section~\ref{sec:setup} we set up our four-dimensional models in a way
suitable for the discussion of moduli stabilization which includes
gaugino condensation and flux quantization. This section is largely
self-contained and the reader mostly interested in the
four-dimensional aspects of our analysis may want to skip
Section~\ref{extension} and move on to Section~\ref{sec:setup}
straight away.  Moduli stabilization within models based on half-flat
mirror manifolds~\cite{GLM} is discussed in
Section~\ref{sec:basicmodels}. In Section~\ref{sec:mixed} we discuss
the models based on the more general half-flat spaces introduced in
Section~\ref{extension}. We conclude in Section~\ref{sec:conclusions}.
Various technical details are deferred to the three appendices.
Appendix~\ref{appA} contains a calculation of the scalar curvature of
the generalised half-flat spaces, which is essential in establishing
the consistency of the generalised models of Section~\ref{extension}.
In Appendix~\ref{appB} we have collected a
number of useful relations on special geometry, while
Appendix~\ref{appC} summarizes our four-dimensional $N=1$ supergravity
conventions. It also includes an elementary proof that supersymmetric
AdS vacua of this theory are always stable.

\section{The heterotic string on half-flat manifolds}
\label{extension}

In this section we will review the compactification of the heterotic
string on half-flat mirror manifolds~\cite{GLM} and present an
extension of this work to the spaces proposed in Ref.~\cite{DAFTV,AT,GLW}.

\subsection{The heterotic string on half-flat mirror manifolds}
\label{sec:hfrev}

Half-flat mirror manifolds arise in the context of type II mirror
symmetry~\cite{GLMW} and can be thought of as mirror duals to
Calabi-Yau manifolds with NSNS flux.  More specifically, given a
mirror pair $(X,Y)$ of Calabi-Yau manifolds, the mirror of, say, IIB
on $Y$ in the presence of NSNS flux is IIA on a half-flat mirror
manifold $\hat{X}$ (without flux). This half-flat mirror manifold
$\hat{X}$ is closely related to the original Calabi-Yau mirror $X$ in
that it can be characterized by the two Hodge numbers $h^{(1,1)}$ and
$h^{(2,1)}$ of $X$ and carries sets of two- three- and four-forms
analogous to the sets of harmonic forms on the Calabi-Yau space $X$.
Specifically, on $\hat{X}$ we denote by $(\omega_i)$ and
$(\tilde{\omega}^j)$ a basis for the two- and four-forms respectively,
where $i,j,\dots = 1,\dots ,h^{(1,1)}$, which satisfy
\begin{equation}
  \label{norm24}
  \int_{\hat X} \ox_i \wg \tox^j = \dx_i^j \; .
\end{equation}
Further, on $\hat{X}$, one can define a set of symplectic three forms
$(\alpha_A,\beta^B)$ where $A,B,\dots = 0,\dots ,h^{(2,1)}$ with
\begin{equation}
  \label{norm3}
  \int_X \ax_A \wg \bx^B = \dx_A^B \; , \quad \int_X \ax_A \wg \ax_B =
  \int_X \bx^A \wg \bx^B = 0 \; .
\end{equation}
Being manifolds with $\mbox{SU}(3)$ structure~\cite{CS}, half-flat
mirror manifolds carry a two-form $J$ and three-form $\Omega$ which
are the analog of the K\"ahler form and the holomorphic $(3,0)$ form
on Calabi-Yau manifolds~\footnote{Although the manifolds discussed in
  this paper are generally neither complex nor K\"ahler, we will
  frequently use Calabi-Yau terminology and, for example, refer to $J$
  as K\"ahler form.}. As on Calabi-Yau manifolds these forms can be
expanded as
\begin{eqnarray}
 J&=&t^i\omega_i \; , \label{J}\\
 \Omega &=&{\cal Z}^A\alpha_a-{\cal G}_A\beta^A\; ,\label{Omega}
\end{eqnarray}
where $t^i$ and ${\cal Z}^A$ are the equivalent of K\"ahler and
complex structure moduli. As usual, the coefficient ${\cal G}_A$ can
be obtained from a holomorphic pre-potential ${\cal G}={\cal G}({\cal
  Z}^A)$, homogeneous of degree two, as
\begin{equation}
 {\cal G}_A=\frac{\partial{\cal G}}{\partial{\cal Z}^A}\; . \label{GA}
\end{equation}
So far, the set-up has been exactly as for Calabi-Yau manifolds. The
main difference is that the forms $(\omega_i)$ and
$(\alpha_A,\beta^A)$ are no longer harmonic but rather satisfy
\begin{equation}
  \label{hfalg}
  d \ox_i = e_i \bx^0 \; , \quad  d \ax_0 = e_i \tox^i \; , \quad  d
  \ax_a = d \bx^A = 0 \; ,  \ \quad d \tox^i = 0 \; .
\end{equation}
Here $e_i$ are $h^{(1,1)}$ parameters (real numbers) which
characterize the torsion of the half-flat mirror manifold under
consideration. 

\vspace{0.4cm}

Having described the basic structure of half-flat mirror manifolds,
let us now review the compactification of the heterotic string (at
lowest order in $\alpha '$) on those spaces. Besides the metric, there
are two other bosonic fields, namely the dilaton $s=\exp (-2\phi )$
and the NSNS two-form $\hat{B}$. The latter can be expanded as
\begin{equation}
  \label{Bexp}
  \hat B = B + \tau^i \ox_i\; ,
\end{equation}
where $B$ is a four-dimensional two-form which can be dualised to a
scalar $\sigma$ and $\tau^i$ is a set of axions. Together with the
dilaton and the K\"ahler moduli, these fields pair up into
four-dimensional chiral multiplets as
\begin{eqnarray}
 T^i &=& \tau^i+it^i \;, \label{Ti}\\
 S &=& \sigma +is\; .\label{S}
\end{eqnarray}
In terms of the projective coordinates ${\cal Z}^A$ the complex
structure chiral multiplets $Z^a$, where $a,b,\dots = 1,\dots
,h^{(2,1)}$, are obtained by $Z^a={\cal Z}^a/{\cal Z}^0$ and we write
these fields as
\begin{equation}
 Z^a=\zeta^a+iz^a\; . \label{Za}
\end{equation}
The K\"ahler potential for those fields in the large radius limit 
is then of the standard Calabi-Yau
form, that is,
\begin{equation}
 K=K^{(T)}+K^{(Z)}+K^{(S)}\; , \label{K}
\end{equation}
with
\begin{eqnarray}
 K^{(T)} &=& -\ln \left( \frac43 \cK \right) \;, \nn \\[2mm]
 K^{(Z)} &=& -\ln \left(\frac43 \tilde{\cK} \right) \;, \label{KS}\\[2mm]
 K^{(S)} &=& -\ln\left(i(\bar{S}-S)\right) \;, \nn
\end{eqnarray}
and
\begin{eqnarray}
 \cK &=& \frac{i}{8}d_{ijk} (T^i - \bar{T}^i) (T^j - \bar{T}^j)
 (T^k - \bar{T}^k) = d_{ijk}t^i t^j t^k \;, \label{k}\\ 
 \tilde{\cK} &=& \frac{3i}{4} \left(\bar{\cal Z}^A {\cal G}_A - 
   {\cal Z}^A \bar{\cal G}_A \right)\; .\label{kt}
\end{eqnarray}
Here, $d_{ijk}$ are numbers analogous to the intersection numbers of the
associated Calabi-Yau space $X$. Later, we will be working in the large complex
structure limit, where the pre-potential $\cG$ can be written as
\begin{equation}
 {\cal G}=-\frac{1}{6} \frac{\tilde{d}_{abc} {\cal Z}^a {\cal Z}^b 
   {\cal Z}^c}{{\cal Z}^0}\; , \label{G}
\end{equation}
with $\tilde{d}_{abc}$ analogous to the intersection of the associated mirror
Calabi-Yau space $Y$. In this case, the complex structure K\"ahler potential
takes a form similar to the one for the K\"ahler moduli, that is
\begin{equation}
 \tilde{\cK} = \frac{i}{8}\tilde{d}_{abc}(Z^a - \bar{Z}^a)
 (Z^b - \bar{Z}^b)(Z^c - \bar{Z}^c)
 =\tilde{d}_{abc}z^a z^b z^c\; . \label{kt1}
\end{equation}

Let us now discuss the superpotential. In Ref.~\cite{GLM} it has been shown
that, for general heterotic compactifications on manifolds with $\mbox{SU}(3)$
structure, the superpotential to order ${\alpha '}$ can be obtained from the 
Gukov-Vafa-Witten type formula
\begin{equation}
 W=\sqrt{8}\int_{\hat{X}}\Omega\wedge (\hat{H}+idJ)\; ,\label{gvw}
\end{equation}
where $\hat{H}$ is the NSNS field strength. For half-flat mirror manifolds this
field strength can be written as
\begin{equation}
  \label{Hhf}
  \hat H = dB + d\tau^i\wg\ox_i + \tau^i e_i \bx^0 + H_{\rm flux} \; .
\end{equation} 
where the first three terms have been computed by taking the
exterior derivative of Eq.~\eqref{Bexp}. Note that the third term is
new and arises because the forms $\omega_i$ are no longer closed, see
Eq.~\eqref{hfalg}. We have also added on an additional NSNS flux
contribution
\begin{equation}
  H_{\rm flux}=(\mu^a\ax_a - \ex_a \bx^a) \; , \label{Hflux}
\end{equation}
with electric and magnetic flux parameters $\ex_a$ and $\mu^a$, respectively.
If we arrange the $\alpha '$ terms in the $\hat{H}$ Bianchi
identity to cancel (for example by choosing the standard embedding)
if follows that $d\hat{H}=0$. For this reason we have dropped the
term proportional to the non-closed form $\alpha_0$ in
Eq.~\eqref{Hflux}. 
We have
also omitted a possible term proportional to $\beta^0$ in \eqref{Hflux} which
can be absorbed into a re-definition of the axions $\tau^i$, as is evident
from Eq.~\eqref{Hhf}. Inserting the field strength~\eqref{Hhf},
the $(3,0)$ form~\eqref{Omega} and $dJ=e_it^i\beta^0$ into the general
formula~\eqref{gvw}, one finds the superpotential
\begin{equation}
  \label{Whf}
  W= \sqrt{8} (e_i T^i + \ex_a Z^a - \mu^a \cG_a) \; ,
\end{equation}
where the basic integrals~\eqref{norm3} have been used.  This result
has been checked in Ref.~\cite{GLM}, where the four-dimensional scalar
potential was calculated from an explicit reduction of the
ten-dimensional bosonic action of the heterotic string. This scalar
potential has three contributions which arise from the third term in
Eq.~\eqref{Hhf} and the NSNS flux~\eqref{Hflux}, both inserted into
the $\hat{H}$ kinetic term, and the non-vanishing scalar curvature of
the half-flat mirror manifolds.  These three contributions lead to a
potential which can be exactly reproduced from the above
superpotential, using the standard relations of four-dimensional $N=1$
supergravity (see Appendix~\ref{appC} for a summary of supergravity
conventions). In the following subsection we will generalize this
calculation to a larger class of manifolds with $\mbox{SU}(3)$
structure.

\subsection{Setup for the extended models}
\label{extmod}

Having discussed the basic models obtained from the compactification
on half-flat mirror manifolds we can now study a generalisation of the
half-flat spaces which was proposed in Ref.~\cite{DAFTV}. The same
class of spaces appeared in \cite{AT,GLW} and it was argued to be the
correct Ansatz for a consistent Kaluza-Klein truncation to four
dimensions. Here we will use the prescriptions given in the above
references, and show that such a truncation is indeed consistent with
supersymmetry. In particular, we will show that the four-dimensional
scalar field potential derived from a compactification on those
generalised spaces is consistent with the Gukov-Vafa-Witten type
formula~\eqref{gvw} for the superpotential.

\vspace{0.4cm}

We start by reviewing the main features of this new class of manifolds
with $\mbox{SU}(3)$ structure which we will denote \emph{generalised
  half-flat manifolds}. We will mostly follow Refs.~\cite{DAFTV,GLW}.
As we have done for the half-flat mirror manifolds, the existence of
two-forms $(\omega_i)$, four-forms $(\tilde{\omega}^i)$ and
three-forms $(\alpha_A,\beta^B)$ satisfying the basic integral
relations~\eqref{norm24} and \eqref{norm3} is postulated.  However,
the crucial differential relations~\eqref{hfalg} are now generalised
to
\begin{equation}
 d \ox_i = p_{Ai} \bx^A - q^A_i \ax_A \; ,\qquad
 d \ax_A  = p_{Ai} \tox^i\; ,\qquad  d \bx^A  = q^A_i \tilde{\omega}^i
 \; ,\qquad d \tox^i=0\; ,\label{extalg}
\end{equation}
with (real) torsion parameters $p_{Ai}$ and $q_i^A$. From $d^2w_i=0$
one concludes that the additional constraints
\begin{equation}
  \label{cons1}
  p_{Ai} q^A_j - q^A_i p_{Aj} = 0 \; ,
\end{equation}
have to be imposed, for a consistent definition of the exterior
derivative. In what follows the above defining relations are enough in
order to derive the low energy action which arises from
compactification on these manifolds. In appendix \ref{appA} we will
have more to say about the geometry of these spaces and, in
particular, about their torsion classes, which differ from those of a
half-flat manifold.

\vspace{0.4cm}

The expansion of the K\"ahler form, $J$, the $(3,0)$ form, $\Omega$,
and the NSNS two-form, $\hat{B}$, in terms of the basic forms remains
unchanged and is given in Eqs~\eqref{J}, \eqref{Omega} and
\eqref{Bexp}. This also means that we have the same set of moduli
fields,\footnote{Strictly speaking the fields we are talking about are
  no longer moduli as the potentials generated are not flat in these
  directions. However we continue to call them moduli in order to
  stress that we are interested in the fields which were the moduli of
  the related Calabi--Yau compactifications.}  namely the K\"ahler and
complex structure moduli $T^i$ and $Z^a$ and the dilaton $S$. Whenever
exterior derivatives are taken we now have to work with the
generalised relations~\eqref{extalg}. This means that the NSNS
three-form field strength associated to~\eqref{Bexp} is given by
\begin{equation}
\hat{H}=dB+d\tau^i\wedge\omega_i+\tau_i(p_{Ai}\beta^A-q_i^A\alpha_a)+H_{\rm
  flux}\; , 
 \label{Hext}
\end{equation}
where, as before, we have added on the NSNS flux part
\begin{equation}
  \label{Hflux1}
  H_{\textrm{flux}} = \mu^A \ax_A - \ex_A \bx^A \; ,
\end{equation}
with electric and magnetic flux parameters $\ex_A$ and $\mu^A$. If the RHS
of the heterotic Bianchi identity
\begin{equation}
 d\hat{H}=\frac{\alpha '}{4}\left(\mbox{tr}(F\wedge
 F)-\mbox{tr}(R\wedge R)\right)\;, \label{BI}
\end{equation}
vanishes (for example, by choosing the standard embedding), then $\hat{H}$
needs to be closed which implies the further constraints
\begin{equation}
  \label{cons2}
  \mu^A p_{Ai} - \ex_A q^A_i = 0 \; ,
\end{equation}
between flux and torsion parameters. On the other hand, the RHS of
Eq.~\eqref{BI}, although necessarily exact, can be non-zero, so that
the constraint~\eqref{cons2} can be avoided by, for example, more
complicated choices of the gauge bundle. It is convenient to
introduce the following combinations
\begin{equation}
  \label{cpxflux}
  \begin{aligned}
    \te_A & = \ex_A - T^i p_{Ai} \;, \\
    \tm^A & = \mu^A -T^i q^A_i \; ,
  \end{aligned}
\end{equation}
of fluxes, torsion parameters and K\"ahler moduli in terms of which the
NSNS field strength can be expressed as
\begin{equation}
  \label{Hint}
  \hat{H} = dB+d\tau^i\omega_i+\Re (\tm^A) \ax_A - \Re (\te_A) \bx^A \; .
\end{equation} 
For the exterior derivative of the K\"ahler form $J$ one finds
\begin{equation}
  dJ = t^id\ox_i = \Im (\tm^A)\ax_A -\Im (\te_A)\bx^A \; ,
\end{equation} 
where the differential relations~\eqref{extalg} and the definitions~\eqref{cpxflux}
have been used. These last two results for $\hat{H}$ and $dJ$, together with
the standard expansion for the $(3,0)$ form~\eqref{Omega} and the
basic integrals~\eqref{norm3}, can be used to evaluate the
formula~\eqref{gvw} for the superpotential. A simple calculation leads to
\begin{equation}
  W = \sqrt 8 ( \te_A Z^A - \tm^A\cG_A) \; .\label{suppot01}
\end{equation} 
We will now verify this result by an explicit reduction of the ten-dimensional
bosonic action.

\subsection{Reduction for the generalised models}
\label{extcalc}

The starting point for the compactification is the lowest order in
$\alpha '$ of the bosonic part of the ten-dimensional effective action
of the heterotic string. This is given by
\begin{equation}
  \label{SB}
  S_{0,{\rm bosonic}} = -\frac{1}{2\kappa_{10}^2} \int_{M_{10}}
  e^{-2 \hat \phi} \left[ \hat R \star \mathbf{1}
    - 4 d \hat \phi \wedge\star  d \hat \phi
    + \frac12 \hat H \wedge \star \hat H \right] \, .
\end{equation}
As the main assumption for compactifications on generalised half-flat
manifolds is that the light spectrum of normal Calabi--Yau (and also
half-flat) compactifications is unchanged, we will not be concerned
with the derivation of the kinetic terms for the various fields one
obtains in four dimensions. They are exactly as discussed for the case
of half-flat mirror manifolds, see Eqs~\eqref{K}--\eqref{kt}.
Instead we concentrate on the scalar potential. As explained in
the previous section, one contribution to
the four-dimensional potential arises from the $\hat{H}$ kinetic term
with \eqref{Hint} inserted. A standard calculation~\cite{TV,JM} leads to
\begin{equation}
  \label{VH}
  e^{-K}V_H = -4e^{-K^{(Z)}} \big[ \Re (\te_A)  - \Re
  (\tm^C) \cM_{CA} \big] \big(\Im \cM^{-1} \big)^{AB} \big[ \Re (\te_A)
  - \Re (\tm^C) \bar \cM_{CA} \big]\; .
\end{equation}
Here the matrix $\cM$ is the period matrix \eqref{Nsg} which, for the
complex structure sector, is also given by the relations \eqref{ABC}.

\vspace{.4cm}

The second contribution arises from the Einstein Hilbert term in~\eqref{SB}
and is due to the non-vanishing scalar curvature of the half-flat spaces.
The calculation of this scalar curvature, for the spaces characterized by
the relations~\eqref{extalg}, is somewhat non-trivial and has been
carried out in Appendix~\ref{appA}. The result is
\begin{eqnarray}
  \label{VR}
  e^{-K}V_R & = & -4e^{-K^{(Z)}} \big[ \Im (\te_A)  - \Im (\tm^C) \cM_{CA}
  \big] \big(\Im \cM^{-1} \big)^{AB} \big[ \Im (\te_A)  - \Im (\tm^C)
  \bar \cM_{CA} \big] \nn \\
  && + 8 E_i {\bar E}_j (g^{ij}-4t^it^j)
  \; ,
\end{eqnarray}
where we have introduced the notation
\begin{equation}
  \label{Ei}
  E_i = p_{Ai} Z^A - q_i^A \cG_A\; .
\end{equation}
It is not hard to see that, provided the constraints \eqref{cons1} and
\eqref{cons2} are satisfied, the total potential takes the form
\begin{equation}
  \label{Vtot}
  e^{-K}V = -4e^{-K^{(Z)}} \big[\te_A  - \tm^C \cM_{CA} \big] \big(\Im
  \cM^{-1} \big)^{AB} \big[ \overline{\te_A  - \tm^C \cM_{CA}} \big] 
  + 8 E_i {\bar E}_j (g^{ij}-4t^it^j) \; .
\end{equation}
We now need to verify that this potential indeed originates from the
superpotential~\eqref{suppot01} via the standard supergravity
formula~\eqref{N1pot}. Since the index $X$ in this formula runs over
all chiral fields which, in our case, consist of the dilaton $S$, the
complex structure moduli $Z^a$ and the K\"ahler moduli $T^i$, we will
discuss each case separately.

\vspace{.4cm}

First of all notice that, since the superpotential \eqref{suppot01}
does not depend on $S$, the contribution of the dilaton-axion
chiral superfield to the potential can be found from \eqref{KS} to be
simply
\begin{equation}
  \label{Spart}
  e^{-K}V_S = |W|^2.
\end{equation}

\vspace{.4cm}
\noindent 
For the complex structure moduli we obtain
\begin{equation}
  \label{DaW}
  \frac{1}{\sqrt 8} D_a W = (\te_B - \cG_{BC} \tm^C) D_a Z^B \; ,
\end{equation}
where we define $\cG_{BC}=\partial_B \partial_C \cG$.
Using the relations \eqref{N2relcs} and \eqref{pM} one immediately finds
\begin{equation}
  g^{a\bar b}D_aWD_{\bar b}\bar W = - 4 
  \cko\big(\Im \cM^{-1} \big)^{AB}
  \left(\te_A - \tm^C \bar\cM_{CA}\right)
  \big(\overline{{\te}_B- {\tm}^D {\bar \cM}_{DB}}\big) 
  - 8 |\te_A \bar{Z}^A - \tm^A \bar{\cG}_A|^2 \label{CSMcont} \; .
\end{equation} 
Note that the first term in the above equation is similar to the first
term of \eqref{Vtot}, except for the complex conjugations which do not
work out quite right. However, it is just a matter of algebra to show
that these two terms are indeed identical provided that the constraints
\eqref{cons1} and \eqref{cons2} hold.

Also, the second term in \eqref{CSMcont} very much resembles the
square of the superpotential, but here the complex
conjugations can not be exchanged so easily. In turn one obtains
\begin{equation}
  \label{CSMcont2}
  g^{a\bar b}D_aWD_{\bar b}\bar W = - 4 
  \cko\big(\Im \cM^{-1} \big)^{AB} \big(\te_A - \tm^C \cM_{CA}\big)
  \big(\overline{\te_B - \tm^D \cM_{DB}}\big)
  -|W|^2 + Y \; ,
\end{equation} 
where by $Y$ we have denoted the combination
\begin{equation}
  \label{Y}
  Y = -32 t^i t^j E_i \bar E_j - 2\sqrt{8} i t^i (E_i \bar W - \bar
  E_i W) \; .
\end{equation}

Let us finally deal with the K\"ahler moduli contribution to the $N=1$
potential.  Using formulae \eqref{Kidef}--\eqref{noscale}
on the K\"ahler moduli space we find
\begin{equation}
  \label{Kpart}
  g^{ij}D_iWD_{\bar j}\bar W =
  g^{ij}\partial_iW\partial_{\bar j}\bar W + 3|W|^2
  +2it^i(W\partial_{\bar i}\bar W-\bar W\partial_iW) \; .
\end{equation} 
To this end it is useful to make the dependence of the superpotential
\eqref{suppot01} on the K\"ahler moduli explicit by writing
\begin{equation}
  W = \sqrt{8} (-E_iT^i + \epsilon_AZ^A-\mu^A\cG_A) \; ,
\end{equation} 
where $E_i$ were defined in \eqref{Ei}. Hence we have
\begin{equation}
  \partial_i W = - \sqrt{8} E_i \; .
\end{equation}
With this we see that the last terms in Eqs~\eqref{Kpart} and
\eqref{Y} cancel identically. Moreover, the $|W|^2$ terms from
equations \eqref{Spart}, \eqref{CSMcont2} and \eqref{Kpart} cancel
against the $-3 |W|^2$ in equation \eqref{N1pot} while the remaining
terms precisely combine into \eqref{Vtot}. This concludes our
derivation of the potential \eqref{Vtot} from the superpotential
\eqref{suppot01}, and establishes a strong argument for the consistency
of the compactifications on the manifolds presented in section
\ref{extmod}, which were introduced in Refs.~\cite{DAFTV,AT,GLW}.

We conclude this section by comparing the superpotential~\eqref{suppot01}
which we have just derived with the one obtained in type IIB compactifications.
There, the fluxes are ``complexified'' in a way that involves the IIB
complex coupling. In our case, the flux parameters are ``complexified''
to $\te$ and $\tm$ in Eq.~\eqref{cpxflux} due to their dependence on the K\"ahler moduli.
Apart from this ``exchange'' of K\"ahler moduli and dilaton, the resemblance
between the two superpotentials is quite striking. This confirms our
expectation that heterotic theories can be as flexible with regard
to moduli stabilization as type II theories when non-trivial torsion
is included.


\section{General structure of low-energy theories} 
\label{sec:setup}

So far we have concentrated on how four-dimensional models arise
from compactifications of the underlying ten-dimensional theory.
In the remainder of the paper we will analyze the implications of
these four-dimensional models for moduli stabilization, and the
purpose of this section is to set up all the necessary ingredients,
in a way that is convenient for this analysis.

\subsection{The models}

From now on, we will adopt the ``phenomenological'' definition of the
chiral superfields in terms of its components, where the real parts
are the ``geometrical'' moduli and the imaginary parts
the axions. With respect to our previous convention, this corresponds to
the simple transformation $\phi^X\rightarrow -i\phi^X$ (together
with a sign flip of the axions) on all fields. Explicitly, this means
we are replacing the field definitions~\eqref{Ti}, \eqref{S} and \eqref{Za}
by
\begin{eqnarray}
  \label{realf}
  S&=&s+i\sigma \; , \\
  T^i&=&t^i+i\tau^i \; , \\
  Z^a&=&z^a+i\zeta^a \; .
\end{eqnarray}
While our general calculation for the four-dimensional effective theory
was valid for all values of the complex structure moduli we will, in the
following, focus on the large complex structure limit. This means that,
from Eqs~\eqref{K}--\eqref{kt1} together with the above field re-definition,
the K\"ahler potential is given by~\footnote{Several numerical factors
  from \eqref{K} to \eqref{kt1} were absorbed into the superpotential
  and the definition of the flux parameters in order to make the
  calculations in the following sections more straightforward.}   
\begin{equation}
 K = -\ln (S+\bar{S}) - \ln (8 \cK) -\ln (8 \tilde{\cK}) \label{Kp}\; ,
\end{equation}
with
\begin{eqnarray}
  \label{KKcs}
  \cK &=& \frac{1}{8}d_{ijk}(T^i+\bar{T}^i)(T^j+\bar{T}^j)(T^k+\bar{T}^k)=
          d_{ijk}t^it^jt^k \; , \\
  \tilde{\cK}&=&\frac{1}{8}\tilde{d}_{abc}(Z^a+\bar{Z}^a)(Z^b+\bar{Z}^b)
                (Z^c+\bar{Z}^c)=\tilde{d}_{abc}z^az^bz^c\; .\label{KKK}
\end{eqnarray}
Recall that $d_{ijk}$ and $\tilde{d}_{abc}$ correspond to the
intersection numbers of the associated Calabi-Yau space and its
mirror, respectively. Both the K\"ahler and complex structure parts of
the K\"ahler potential are given in terms of special geometry
pre-potentials which, due to large radius and complex structure, are
determined by cubic polynomials. The cubic nature of the
pre-potentials means both moduli spaces constitute examples of very
special geometry. Some useful relations for very special geometry,
which we will apply subsequently, are collected in Appendix~\ref{appB}.

\vspace{0.4cm}

Let us now turn to the superpotential. Inserting the
derivatives~\eqref{GA} of the large-complex structure
pre-potential~\eqref{G} into Eq~\eqref{suppot01}, along
with the definitions~\eqref{cpxflux} of the complex flux parameters, the
explicit form of the superpotential ${\cal W}$ turns out to be
\begin{eqnarray}\label{Wexp}
    {\cal W} &=& -i(\ex_0 - i T^i p_{0i}) + (\ex_a - i T^i p_{ai}) Z^a +
  \frac{i}{2} (\mu^a - i T^i q^a_i) \tilde d_{abc} Z^b Z^c\nonumber\\
             && + \frac16(\mu^0 - i T^i q^0_i) \tilde d_{abc} Z^a Z^b
  Z^c\; .
\end{eqnarray}
As we have pointed out in section~\ref{extmod}, the parameters in 
this superpotential are not
independent but satisfy
\begin{eqnarray}
        p_{Ai} q^A_j - q^A_i p_{Aj} & = & 0 \;, \label{c1}\\
        \ex_A q^A_i - \mu^A p_{Ai}  & = & 0 \; . \label{c2}
\end{eqnarray}
Note that the first of these constraints follows from the property $d^2=0$
of the exterior derivative and is, therefore, strictly necessary. The
second one is a consequence of $d\hat{H}=0$, which is the correct form
of the heterotic Bianchi identity if the $\alpha '$ corrections on the
RHS of Eq.~\eqref{BI} cancel by themselves, for example, by choosing
the standard embedding. However this need not be the case,
so that this second constraint can be avoided.\footnote{Since the
calculation in section \ref{extcalc} relies on equation \eqref{c2} we
may argue that this constraint cannot be relaxed. However, if
we were to incorporate consistently all the terms which appear at
order $\ax'$, we would expect to find the same superpotential as
before. In fact, this is precisely what the Gukov-Vafa-Witten
formula~\eqref{gvw} evaluated for a field strength $H$ which includes
the $\alpha '$ corrections predicts.}  In this paper, we will study
both cases with and without the second constraint. 
Finally, note that half-flat mirror
manifolds correspond to the special case where we set
$\epsilon_0=p_{ai}=q_i^a=\mu^0=q_i^0=0$ and $p_{0i}=-e_i$ in the
superpotential~\eqref{Wexp}. This leads to
\begin{equation}
 {\cal W}=e_iT^i+\epsilon_aZ^a+\frac{i}{2}\tilde{d}_{abc}\mu^aZ^bZ^c
 \;, \label{Whf1}
\end{equation}
which is the large complex structure limit of Eq.~\eqref{Whf}, as it
should. 

The above K\"ahler potential and superpotential feed into the general
formula for the four-dimensional $N=1$ supergravity potential and we
have summarized the relevant conventions in Appendix~\ref{appC}. In
this paper, we will only be concerned with supersymmetric vacua of
these potentials, that is, solutions to the F-equations. Generically,
such solutions have a negative cosmological constant~\eqref{V0} and so
they lead to four-dimensional AdS vacua. It is known~\cite{DNP} that
such vacua are always stable and Appendix~\ref{appC} also contains an
elementary proof of this fact.

\subsection{Gaugino condensation}

As the dilaton does not appear in the superpotential ${\cal W}$,
Eq.~\eqref{Wexp}, the potential will usually be runaway in this
direction. Hence, if we want to have any chance of stabilizing all
moduli, we should consider additional contributions.  As has been
shown in Ref.~\cite{GLM}, the gauge kinetic function $f$ of the
four-dimensional gauge group ${\rm SO}(10)\otimes E_8$ for heterotic
compactifications on half-flat mirror manifolds is given by
\begin{equation}
 f=S \; ,
\end{equation}
to leading order. Clearly this result extends to the generalised
half-flat manifolds discussed in the previous section and more general
gauge bundles.
Hence, hidden-sector gaugino condensation~\cite{DRSW} 
leads to an additional dilaton-dependent superpotential term which 
is precisely what we need. We will, therefore consider the superpotential
\begin{equation}
 W={\cal W}+ke^{-cS} \; ,\label{Wgau}
\end{equation}
with ${\cal W}$ as given in Eq.~\eqref{Wexp}. Here $k$ and $c$ are
constants, the latter being determined by the one-loop beta function
of the gauge group. To make this more precise, we normalize the real
part of the dilaton, $s$, such that
\begin{equation}
 s= \frac{1}{\alpha_{\rm YM}}=\frac{4\pi}{g_{\rm YM}^2}\; ,
\end{equation}
where $g_{\rm YM}$ is the Yang-Mills coupling constant. In terms of
the one-loop beta function coefficient $b$, the constant $c$ can then
be written as $c=6\pi /b$. For gauge group $E_8$, one finds $b=90$
and, hence,
\begin{equation}
 c=\frac{\pi}{15}\; .\label{c}
\end{equation}
The pre-factor $k$ is hard to fix precisely not least because
corrections due to the two-loop beta-function will lead to an
$S$-dependent pre-factor of the exponent in $W$, which we neglect
in the present context. We will simply parameterize $k$ as
\begin{equation}
  \label{kval}
  k = \frac{\tilde k}{\sqrt{\ax'}} \;,
\end{equation}
where $\alpha '$ is the string tension and $\tilde k$ is a
dimensionless constant which one expects to be of order one.

\subsection{Quantization of flux and torsion}
\label{sec:quant}

We would now like to be somewhat more specific about the
quantization of flux and torsion parameters in the superpotential. For
the genuine fluxes this is easy to achieve~\cite{RW} by imposing that $H$
is an element in the integral cohomology (modulo normalization factors). It is
less straightforward to see how the torsion parameters of the internal
manifold should be quantized. For the half-flat mirror manifolds, this
will be done via the mirror symmetry relation which was used in order
to establish the existence of such spaces in the first place.
Unfortunately, such a correspondence is not known for the more
general manifolds described in section \ref{extmod}, so we will have
to make a plausible assumption about quantization for these spaces,
generalizing from the results obtained for half-flat mirror manifolds.

Before we can find the quantization rules for the flux parameters we
should fix the normalization of our moduli fields. We recall that the
above models have been derived and are valid in the large radius and
large complex structure limit. Hence, we adopt a normalization of
fields where these limits correspond to field values
\begin{equation}
 t^i>1\; ,\qquad z^a>1\; .
\end{equation}
What does this convention imply for the underlying internal geometry?
Recall that the dimensionless K\"ahler moduli fields measure
the volume of the various Calabi-Yau two-cycles $C_2^i$ in
units of some (six-dimensional) reference volume $v$. More precisely,
we have
\begin{equation}
  \label{tdef}
  t^i = \frac{1}{v^{1/3}} \int_{C_2^i} J \; ,
\end{equation}
where $J$ is the Calabi-Yau K\"ahler form.
In order to assure that $t^i>1$ indeed corresponds to the limit in which the
``radius'' of these cycles is bigger than one in string units, one has to
fix this reference volume to be\footnote{Of course, there is always an
ambiguity of factors of $2 \pi$ in this calculation which cannot be easily
fixed. To arrive at the result~\eqref{vscale}, we have used two-tori which
should lead to a conservative bound on $t^i$.}
\begin{equation}
  \label{vscale}
  v = (4 \pi^2 \ax')^3 \; .
\end{equation}

In order to fix the normalization of the complex structure moduli in a
similar way, it is useful to consider the mirror picture. The fields
$z^a$ measure the size of two-cycles on the mirror and large radii for
these two-cycles corresponds to large complex structure in the
original model. The volume of these mirror two-cycles should be
measured in units of the same reference volume~\eqref{vscale}, in
order for the two four-dimensional effective theories from the
original model and its mirror to be identical (and the mirror map
being trivial on the four-dimensional fields). With this convention,
it is then clear that $z^a>1$ indeed corresponds to the large complex
structure limit.

We are now ready to discuss the quantization of the flux
parameters which appear in the superpotential~\eqref{Whf}. 
The basic quantization condition for the NSNS three-form field strength
$H$ is given by~\cite{RW}
\begin{equation}
  \label{Hquant}
  \int_{C_3} H = 4 \pi^2 \alpha' \cdot l \;,
\end{equation}
where $l$ is an integer and $C_3$ represents any three-cycle in the
integer homology. Following the explicit calculation of the four-dimensional
potential by dimensional reduction in Ref.~\cite{GLM}, it is easy to see that
this quantization rule, together with the scale convention~\eqref{vscale},
implies that
\begin{eqnarray}
  \label{epsmuquant}
    \ex_a & = & \frac{6 \sqrt 2}{\pi \sqrt{\ax'}} \te_a \; , \\
    \mu^a & = & \frac{6 \sqrt 2}{\pi \sqrt{\ax'}} \tm^a \; ,
\end{eqnarray}
where $\te_a$ and $\tm^a$ are integers. Note that the counterintuitive
numerical factors include the redefinitions of the flux parameters,
which were needed in order to rewrite the K\"ahler potential and
superpotential in the simpler form of \eqref{Kp} to \eqref{Wexp}. 
In order to fix the quantization of the
electric torsion parameters, $e_i$, we should again consider mirror
symmetry.  On the mirror, these electric torsion parameters become
electric flux parameters of the NSNS form. Given that our basic choice
of unit is given by $v$ in Eq.~\eqref{vscale}, both on the original
space and on the mirror, the parameters $e_i$ are quantized in
precisely the same way as $\ex_a$ and $\mu^a$, that is,
\begin{equation}
  \label{equant}
  e_i = \frac{6 \sqrt 2}{\pi \sqrt{\ax'}} \tilde e_i \; , 
\end{equation}
where $\tilde{e}_i$ are integers.

Finally let us comment on the other parameters which will appear in
our discussion and that we did not discuss here. Given the above
conventions all the flux/torsion parameters are quantized in terms of
the same unit, and we shall assume the same for the
flux~\footnote{The quantization of flux parameters in the
generalised half-flat models can be discussed in more detail by
studying their third cohomology and homology. It is
likely to be more subtle than assumed in this paper.} and torsion
parameters of the more general models considered in section
\ref{extmod}. This is far from being a rigorous treatment, but
the most natural and straightforward assumption one can make in the
absence of detailed knowledge about these manifolds.

\section{Vacua of the basic models}
\label{sec:basicmodels}

In this section we study moduli stabilization for the
four-dimensional model based on half-flat mirror manifolds, as introduced
in section \ref{sec:hfrev}. For clarity
we start with a simplified version where we consider only one size
modulus, $T$, and one shape modulus, $Z$, together with the axio-dilaton,
$S$. Later on in this section we will generalize our discussion to
arbitrary numbers of $T$ and $Z$ moduli. Throughout, we will focus
on supersymmetric solutions of the above systems.

\subsection{The STZ model}

For the simple three-field model with one K\"ahler modulus $T=t+i\tau$, one complex
structure modulus $Z=z+i\zeta$ and the dilaton $S=s+i\sigma$, the K\"ahler potential
\eqref{Kp}--\eqref{KKK} specializes to
\begin{equation}
  \label{simpleKp}
   K = - \ln{(S + \bar S)} - 3 \ln{(T + \bar T)}- 3 \ln{(Z + \bar Z)}\; ,
  \end{equation}
where we have set $d_{111}=\tilde d_{111}=1$.  The flux/torsion
superpotential \eqref{Whf1} now simply reads
\begin{equation}
  \label{ssuppot}
  {\cal W} = eT + \epsilon Z + \frac{i\mu}{2}Z^2 \;,
\end{equation} 
and, including the gaugino condensate term, we have
\begin{equation}
 \label{WSTZ}
 W={\cal W}+ke^{-cS}\; .
\end{equation}
The F-equations for this model become
\begin{eqnarray}
  F_T & = & e -\frac{3}{2t}W = 0 \; , \label{FT}\\
  F_Z & = & \epsilon +i\mu Z-\frac{3}{2z}W = 0\; , \label{FZ}\\
  F_S & = & -kce^{-cS} -\frac{1}{2s}W = 0 \;  \label{FS}.
\end{eqnarray} 
The solution to \eqref{FT} implies that 
\begin{equation}
 W = \frac{2e}{3}t \;, \label{Wval}
\end{equation}
which is a real quantity. Inserting this into \eqref{FZ} we find
\begin{equation}
  \label{FZres}
  \begin{aligned}
    \epsilon z & = et \; ,\\
    \mu z & = 0 \; .
  \end{aligned}
\end{equation} 
Recall that our model is valid only in the regime of large
volume and complex structure and, in particular, we have $t,z \ne 0$.
Therefore, the second of equations \eqref{FZres} implies the vanishing
of the magnetic flux term, that is $\mu=0$. Let us absorb the 
constant $c$ in the gaugino condensate potential by defining the quantities
\begin{equation} 
 x=cs\; ,\qquad y=c\sigma\; .
\end{equation}
Then, using \eqref{Wval}, Eq.~\eqref{FS} can be written as
\begin{equation}
  \label{FSres}
  \begin{aligned}
    -2kxe^{-x} \cos{y} & = \frac{2et}{3} \;, \\
    2kxe^{-x} \sin{y} & =  0 \; .
  \end{aligned}
\end{equation} 
The value $x=0$ is unacceptable, as it would correspond to the strong
(gauge) coupling limit. Consequently we have to impose $\sin{y} = 0$
which fixes $y$ to $y=n\pi$ for some integer $n$.\footnote{Note that
had we taken no gaugino condensate, that is $k=0$, the above system admits
a solution only if $e=\ex=\mu=0$. This is the limit of compactifying
on a normal Calabi--Yau manifold with no fluxes turned on and it is in
agreement with the result derived in Ref. \cite{AS1} that the internal
manifold has to be complex in order to obtain supersymmetric
solutions.}  Finally, calculating directly the real and imaginary parts
of $W$ in Eq.~\eqref{WSTZ} by inserting~\eqref{FZres}, $y=n\pi$  and
$\mu =0$, we can evaluate the constraint~\eqref{Wval}. Combining all
results we find the most general supersymmetric solution of our model to be
\begin{eqnarray}
  \label{values}
  e\tau & = & -\epsilon\zeta \; , \nn \\
  et & = & \epsilon z = (-1)^{n+1}\frac{3k}{4}e^{-1/4} \;, \nn \\
  x & = & \frac{1}{4}\;, \\
  y &=& n\pi\; . \nn 
\end{eqnarray} 
Let us discuss this result. These equations fix $t$, $z$, $s$ and
$\sigma$, and the same
holds for $e \tau + \epsilon \zeta$, while the orthogonal combination
remains a flat direction. It is clear from the above
expressions that, in order to be in the large radius and complex
structure limits, the torsion and flux parameters $e$ and $\epsilon$
should be sufficiently small. However, as those parameters are
quantized, the best we can do is to stick to their minimal,
non-vanishing, values which
corresponds to $|\tilde{e}|=|\tilde{\epsilon}|=1$ in equations
\eqref{epsmuquant}. Even for this
choice, we need a value of $\tilde k$ bigger than $5$ to arrive
at $t>1$ and $z>1$. In other words, it is difficult to stabilize
fields in the large radius and large complex structure region and,
only by going to the limit of what one would consider reasonable parameter
choices, can marginally consistent solutions be obtained.

There is a similar problem with the gauge coupling since $x$ is fixed
at a relatively small value for the above solutions. Even using
the relatively large $E_8$ beta-functions coefficient~\eqref{c} we find
for the inverse gauge coupling
\begin{equation}
  s = x/c = \frac{15}{4\pi} \sim 1.19 \;,
\end{equation} 
which is barely in the weak coupling limit. 

\subsection{The general case}

Let us now briefly discuss the general case, where $h^{(1,1)}$ and
$h^{(2,1)}$ are arbitrary integers. With the K\"ahler potential
as in Eqs~\eqref{Kp}--\eqref{KKK} and the superpotential~\eqref{Wgau}, \eqref{Whf1}, we derive the following F-equations
\begin{eqnarray}
  F_{T^i} & = & e_i +K_iW = 0 \;, \label{FT2}\\
  F_{Z^a} & = & \epsilon_a +i\mu^bd_{abc}Z^c +K_aW = 0 \;, \label{FZ2}\\
  F_S & = & -kce^{-cS} -\frac{1}{2s}W = 0\label{FS2}\; ,
\end{eqnarray} 
with $K_i$ and $K_a$ given by
\begin{equation}
  \label{firstK}
  K_i = -\frac{3}{2\cK}d_{ijk}t^jt^k\; ,
  \quad K_a = -\frac{3}{2\tilde\cK}d_{abc}z^bz^c \; .
\end{equation}
Note that $K_i$ and $K_a$ and, hence, $W$ are real with
the latter given by 
\begin{equation}
  W= \frac{2e_it^i}{3} \;.\label{ReW}
\end{equation} 
As a consequence, taking the imaginary part of Eq.~\eqref{FZ2} gives
\begin{equation}
  d_{abc}z^b\mu^c = 0 \;.
\end{equation} 
The matrix $\cK_{ab} = d_{abc}z^c$ is non-singular for a physical
point $z^a$ in moduli space (as otherwise the K\"ahler metric $K_{ab}$
would be singular at this point), so all magnetic fluxes must
vanish in order to have a supersymmetric solution. Equation \eqref{FS2}
reproduces similar results to the case with only one $T$ and one $Z$, namely
$y$ is constrained to take the values $y=n\pi$ with $n$ integer, while $x$
must obey 
\begin{equation}
  (-1)^{n+1} 2kxe^{-x} = \frac{2e_it^i}{3} \;.
\end{equation} 
As before, we can compute the value of the superpotential directly by inserting
$\mu^a=0$, $y=n\pi$ and Eq.~\eqref{FZ2} into Eq.~\eqref{Whf1}. On the other hand,
we know from Eq.~\eqref{ReW} that the imaginary part of $W$ must vanish which
leads to the constraint
\begin{equation}
  e_i\tau^i + \epsilon_a\zeta^a = 0\; .
\end{equation} 
This will be the only relation involving the axions, so we can
only fix one of them while we are left with $h^{(1,1)} + h^{(2,1)} -1$
flat axion directions. Matching the real part of $W$ with Eq.~\eqref{ReW}
fixes the value of the dilaton to
\begin{equation}
  x = \frac{1}{4}\; ,
\end{equation} 
while the $t^i$ and $z^a$ moduli obey
\begin{equation}
  e_i = (-)^{n+1}\frac{3k}{4}e^{-1/4} \frac{\di_{ijk}t^j t^k}{\cK}\quad ,
  \quad \epsilon_a = (-)^{n+1}\frac{3k}{4}e^{-1/4} \frac{\tilde\di_{abc}z^b z^c}{\tilde\cK} \; . 
\label{soluce}\end{equation} 
It appears that generic analytic solutions to these last equations for
$t^i$ and $z^a$ cannot be written down but, of course, solutions can
be obtained, either analytically in simple cases or numerically, once
explicit sets of intersection numbers $d_{ijk}$ and $\tilde{d}_{abc}$
have been fixed in the context of a particular model. We will not carry this out
explicitly, as we have already seen that there exist flat axion
directions and that the value of the dilaton is unchanged from the simple
three-field case, so that weak gauge coupling is difficult to achieve.
However, it is clear that solutions to Eqs~\eqref{soluce} will be of
the form $t^i\sim k/(\mbox{flux or torsion})$ (and similarly for
$z^a$) so that flux/torsion quantization makes it hard to obtain vacua
in the large radius and large complex structure limits. In summary,
we have seen that the general model shows all the major problems that
we have already found in the simple three-field case.

\vspace{0.4cm}

Let us consider if there are any alternative 
ways around the above problems. Clearly, some of the difficulties
arise because the supersymmetry condition forces us to set the
magnetic fluxes $\mu^a$ to zero. This problem may not arise for
non-supersymmetric vacua. However, we note that the scalar
potential only depends on the combination $e_i\tau^i$ of the axions $\tau^i$
(since this is true for the superpotential and the K\"ahler
potential is axion independent). Hence, even for non-supersymmetric
vacua we will have at least $h^{1,1}-1$ flat directions. A possible
way forward could then be to study non-supersymmetric solutions
for models with only one $T$ modulus, that is, $h^{1,1}=1$. We will
not do this in the present paper, as we focus on supersymmetric vacua,
but we simply note that, for models with $h^{1,1}=1$, there is still
a chance for consistent (non-supersymmetric) vacua with all moduli fixed.

Another possibility is to modify the superpotential~\eqref{Whf1} to
\begin{equation}
  {\cal W}=e_iT^i+\frac{i}{2}d_{ijk}m^iT^jT^k+\epsilon_aZ^a+\frac{i}{2}\tilde{d}_{abc}\mu^aZ^bZ^c\; ,
      \label{Whf2}
\end{equation}
that is, by including magnetic torsion terms with torsion parameters $m^i$.
Although this is a suggestive extension of the basic model, with a
superpotential perfectly symmetric between the K\"ahler and complex
structure moduli parts, we do not currently know of a convincing
derivation of such a model in the context of the heterotic string.
Given this situation, we will only give a very brief summary
of the results for moduli stabilization we have obtained for such
models. We find that there exist supersymmetric vacua with
all moduli stabilised and values of the dilaton $x$ in the range
$x\in [0,1]$. For suitable choices of parameters $x\simeq 1$ can
be achieved and, with the $E_8$ beta-function coefficient~\eqref{c},
this implies an inverse gauge coupling of at most $s\simeq 4.8$. This
is in the weak coupling region, although still well away from
the ``phenomenological'' value $s\simeq 24$. The values of $t^i$ and
$z^a$ are proportional to the magnetic torsion/flux parameters,
that is, $t^i\sim m^i$ and $z^a\sim\mu^a$, but the constant
of proportionality in these relations is such that large radius
and large complex structure can barely be achieved by minimal
flux/torsion parameters and a value of $k$ at the upper end of
the reasonable range. In summary, adding a magnetic torsion term
can solve two of the three problems of the basic model, namely
fix all moduli and generate weak coupling (although perhaps not
to the desired extent), but achieving large radius and large complex
structure remains problematic.

\vspace{0.4cm}

Why is it so difficult to generate sufficiently large values of $t^i$
and $z^a$? In all examples the values of these fields were basically
determined by an expression of the form $k/(\mbox{flux or
torsion})$. The lower bound on the flux/torsion parameters due to
quantization, combined with the fact that $k$ is expected to be not too
large in $\alpha '$ units, rules out large field values. The 
proportionality of the field values to the constant $k$
in the gaugino condensate potential can be traced back to the fact
that the flux/torsion part ${\cal W}$ of the superpotential does not
have non-trivial globally supersymmetric solutions by itself. For
example, from the superpotential~\eqref{Whf1} for the basic model we
have $W_{T^i}=e_i$ which has no solution unless the torsion parameters
$e_i$ vanish. On the other hand, if ${\cal W}$ had globally supersymmetric
solutions, the values of $t^i$ and $z^a$ at this global level would be
determined by fluxes and torsion parameters only. Provided the locally
supersymmetric solution can be obtained as a perturbation of the
global one this would essentially decouple the values of $t^i$ and
$z^a$ from $k$ and potentially solve our problem. To understand the
local solution as a perturbation of the global one, the value ${\cal
W}_0$ of the superpotential taken at the global solution should be
small. A small $|{\cal W}_0|$ also facilitates weak coupling as
should be intuitively clear from the structure of the
superpotential~\eqref{Wgau}. We will explain those statements in more
detail in the following section, where we analyze the models based on 
generalised half-flat spaces. As we will see, for those models
${\cal W}$ has, in general, global supersymmetric solutions and,
under certain conditions, $|{\cal W}_0|$ can indeed be made small.

\section{Generalised half-flat models}
\label{sec:mixed}

In this section, we will analyze moduli stabilization for the
generalised half-flat models introduced in Section~\ref{extmod}. They
are significantly more complicated than the basic models of the
previous section as they involve more flux/torsion parameters per
field. It will therefore be harder to find simple analytic solutions
and we will have to use approximations and numerical methods.
Also, the main part of our discussion will be within the simplest
three-field STZ model, where $h^{(1,1)}=h^{(2,1)}=1$. This model
already contains eight flux and torsion parameters. However, our main
results should carry over to the general case, which we will discuss at
the end of the section.

\subsection{Relation between locally and globally supersymmetric solutions}

Before we launch into the analysis of the STZ model, we would like 
to understand the relation between globally and locally supersymmetric solutions
of our models in general. This will also provide us with a practical way
of finding supersymmetric vacua. We start with a superpotential of the form
\begin{equation}
 W={\cal W}+ke^{-cS} \;,
\end{equation}
where ${\cal W}={\cal W}(T^i,Z^a)$ is independent of the dilaton and
stands for the flux/torsion part of $W$. For the purpose of the
present discussion we can keep ${\cal W}$ arbitrary but, of course,
we have in mind the concrete form~\eqref{Wexp}. Let us assume that
$T^i=T^i_0$ and $Z^a=Z^a_0$ is a globally supersymmetric minimum,
that is, it satisfies $\partial_i {\cal W}(T_0,Z_0) = \partial_a{\cal W}
(T_0,Z_0)=0$, and let ${\cal W}_0$ be the value of the superpotential
at this minimum 
\begin{equation}
 {\cal W}_0={\cal W}(T_0,Z_0)\; .
\end{equation}
Further, let $M_0$ be the typical moduli mass at this minimum,
computed at the global level from the second derivative of ${\cal
W}$ and let us assume that $|{\cal W}_0/M_0|\ll 1$. It is not hard to
see that this condition is sufficient to ensure that the F-equations,
$F_i=0$ and $F_a=0$, are approximately satisfied by the global
solutions $T^i_0$ and $Z^a_0$, up to small corrections 
\begin{equation}
  \label{dxTZ}
  \dx T^i \simeq \dx Z^a \simeq \frac{\cW_0}{M_0} \; .
\end{equation}
Note that, in the above analysis, we have used the fact that we are working in
the large radius and large complex structure limits, that is, the moduli fields $t^i$
and $z^a$ are bigger than one. Values much larger than one for
these fields will make the approximation even better. For smaller field
values (for example near conifold points) the above argument would have
to be refined.

The specific flux/torsion superpotentials and their derivatives will
generically be of order one or larger due to the quantization of flux
and torsion parameters. To satisfy the above condition it will,
therefore, be sufficient to have $|{\cal W}_0|\ll 1$ in $\alpha '$ units.

Let us turn now to the dilaton F-equation
\begin{equation}
  F_S= \partial_S \cW + K_S \cW = 0 \; .
\end{equation}
Expanding $\cW$ around the global solution and using \eqref{dxTZ}, the
leading contribution to the above equation will come from
\begin{equation}
 F_S\simeq -cke^{-cS}-\frac{1}{2s}\left({\cal W}_0+ke^{-cS}\right)=0 \;,
\end{equation}
which then yields the solution
\begin{equation}
  \label{xsol}
  \begin{aligned}
    (2x+1)e^{-x} = \left|\frac{{\cal W}_0}{k}\right| \;, \\[2mm]
    y =-\mbox{arg}\left(-\frac{{\cal W}_0}{k}\right) \;,
  \end{aligned}
\end{equation}
for the rescaled dilaton
\begin{equation}
 x=cs\; ,\qquad y=c\sigma\; .
\end{equation}
It is this class of supersymmetric solutions which we will be looking
for in our generalised models. For such vacua the values of $T^i$ and $Z^a$
are determined at the global level, thereby potentially circumventing
the problems of achieving large radius and large complex structure
encountered in the previous section. Furthermore, it is clear from Eq.~\eqref{xsol}
that a small value $|{\cal W}_0|$ facilitates a large value of the 
dilaton $x$ and, hence, weak coupling. 

Note that 
this procedure is slightly different to the one outlined in
Refs~\cite{dA1,dA2}, where the issue of integrating out heavy fields
in SUSY theories was addressed and applied to the
KKLT~\cite{KKLT} scenario. Whereas in most papers the
F-equations are used to integrate out heavy moduli we, as indicated above, start
with a globally supersymmetric solution and, by making $|{\cal W}_0|$ small,
ensure that an approximate solution to the full F-equations exists.
In fact, we have numerically checked our procedure and verified that
a solutions of the full F-equations indeed exist close to the globally
supersymmetric ones, provided $|{\cal W}_0|$ is small. An explicit example
will be presented in the next subsection.   

\subsection{The STZ model}

In this subsection, we discuss the three-field model with a single
K\"ahler modulus $T$, a single complex structure modulus $Z$ and
the dilaton $S$.  From Eqs~\eqref{Kp}--\eqref{KKK} the K\"ahler
potential reads
\begin{equation}
  \label{simpleKp1}
   K = - \ln{(S + \bar S)} - 3 \ln{(T + \bar T)}- 3 \ln{(Z + \bar Z)}\; .
\end{equation}
The general flux/torsion superpotential~\eqref{Wexp} becomes
\begin{equation}
  \label{Wmix}
  {\cal W} = i(\xi + i e T) + (\ex + i p T) Z + \frac{i}{2} (\mu + i q
  T) Z^2 + \frac16 (\rho + i r T) Z^3 \; ,
\end{equation}
where we have chosen the signs of the flux parameters $\xi$, $\ex$,
$\mu$ and $\rho$ and the torsion parameters $e$, $p$, $q$ and $r$ for
convenience.  We recall those parameters are subject to a number of
constraints~\eqref{c1} and \eqref{c2}. However, the first set of these
constraints~\eqref{c1} is trivially satisfied for $h^{(1,1)}=1$ while
the second set reduces to the single condition
\begin{equation}
  \label{simplecons}
  \xi r - \ex q + \mu p - \rho e = 0\; .
\end{equation}
We remind the reader that this constraint originates from the relation
$d\hat{H}=0$, which is a consequence of the Bianchi
identity~\eqref{BI} if the order $\alpha '$ terms on the RHS
cancel. This happens, for example, if the standard embedding is chosen,
that is, if the gauge connection is set equal to the spin connection.
On the other hand, for more general gauge bundles the constraint can
be avoided. We will discuss both cases, with and without the
constraint~\eqref{simplecons}. As usual, we take the full superpotential
to be
\begin{equation}
 W={\cal W}+ke^{-cS} \;,
\end{equation}
with ${\cal W}$ as in Eq.~\eqref{Wmix}.

\vspace{0.4cm}

Following the procedure outlined at the beginning of the section,
we will start by searching for global supersymmetric vacua
of ${\cal W}$. These can be found from
\begin{equation}
  \begin{aligned}
    {\cal W}_T =& -e+ipZ-\frac{q}{2}Z^2+i\frac{r}{6}Z^3=0 \;,\\
    {\cal W}_Z =& \epsilon +ipT+(i\mu -qT)Z+\frac{1}{2}(\rho +irT)Z^2=0\; .
  \end{aligned}
\end{equation}
For $r\neq 0$, the first of these equations is a cubic in $Z$ which can be
explicitly solved using Cardano's formula. One finds that for each
choice of the flux/torsion parameters there is
exactly one solution with $z>0$ if and only if the discriminant
of the cubic is positive. The second equation can then be
solved for $T$ in terms of $Z$. 

For $r=0$ the solutions to the previous equations take the simple form
\begin{eqnarray}
 z^2 &=&-\frac{2e}{q}-\frac{p^2}{q^2} \;,\nn \\
 \zeta &=& \frac{p}{q} \;, \label{r0} \\
 tz &=& \frac{\epsilon}{q} - \frac{\mu p}{q^2} - \frac{\rho e}{q^2}
 - \frac{\rho p^2}{q^3} \;, \nn \\
 \tau &=& \frac{\mu}{q}+\frac{\rho p}{q^2}\; . \nn
\end{eqnarray}
For a given set of parameters we can now compute ${\cal W}_0$ and
check whether it is much smaller than one. However, the size of the 
flux parameter space is such that it is virtually impossible to carry
out an analytic search of favoured regions. Therefore we resort to a 
numerical scan, varying the flux/torsion values (which are integers) 
in a certain range from $-M,\dots ,M$. 

We have found the
corresponding vacuum solution for each set of parameters, keeping
only those vacua in the large radius and large complex structure limits,
that is, with $t>1$ and $z>1$. We first carried out this
procedure imposing the constraint~\eqref{simplecons}. For the case
$r\neq 0$ we find, for flux/torsion parameters in the range $-20,\dots ,20$,
that $|{\cal W}_0|>0.8$ always. A similar lower bound in $|{\cal W}_0|$
is found for $r=0$ where we have searched in the range $-70,\dots ,70$.
Furthermore, the lower bound for $|{\cal W}_0|$ is reached for relatively
small values for the flux/torsion parameters and $|{\cal W}_0|$ does
not decrease any further as the parameters range is increased.
We take these results as strong evidence that small values of
$|{\cal W}_0|$ cannot be obtained within this model and, hence,
that it will be difficult to achieve weak coupling.

\vspace{0.4cm}

Next, we have repeated the above procedure but without imposing the
constraint~\eqref{simplecons} focusing, for simplicity, on the case
$r=0$. The results are surprisingly different from the ones obtained
when the constraint is imposed. In particular, we find that vacua with
small values of $|{\cal W}_0|$ can be found without any problem once
the constraint is dropped. A useful way of summarizing the result is
to introduce the quantity $N=N(M,w)$, defined as the number of 
vacua (with $t>1$ and $z>1$) found in the range $-M,\dots ,M$ of flux/torsion
parameters, with associated superpotential values ${\cal W}_0$
satisfying $|{\cal W}_0|< w$.  Numerically, we find that $N$ is
well-described by the scaling law
\begin{equation}
 \label{Neq}
 N(M,w)=N_0M^x w^y \;,
\end{equation}
where $N_0$, $x$ and $y$ are real constants. From a numerical search
of all flux/torsion parameters with $M=10,20,\dots ,70$ we find that
\begin{equation}
 N_0\simeq 0.17\; ,\qquad x\simeq 5.0\; ,\qquad y\simeq 2.1\; .
\end{equation}
This can be easily seen from the following two figures. 
\begin{figure*}[!htb]
\begin{center}
\includegraphics[width=10cm]{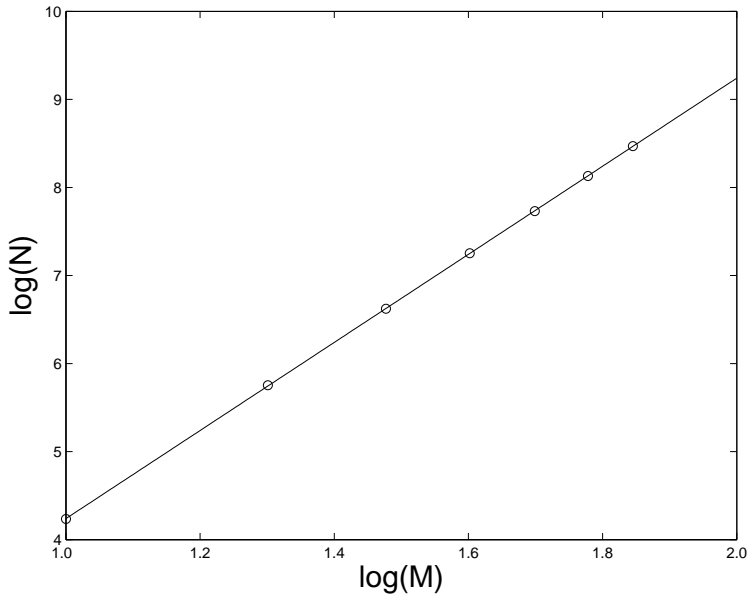}
\caption[fields]{\label{fig0} Total number of vacua, $N$,
as a function of the range of flux/torsion parameters, $M$ (in
logarithmic units).}
\end{center}
\end{figure*}
In Figure~\ref{fig0} we have plotted the total number of solutions with
$|W_0| <1$ (i.e. taking $w=1$ in equation \eqref{Neq}) for
$M=10,20,\dots ,70$.
\begin{figure*}[!htb]
\begin{center}
\includegraphics[width=10cm]{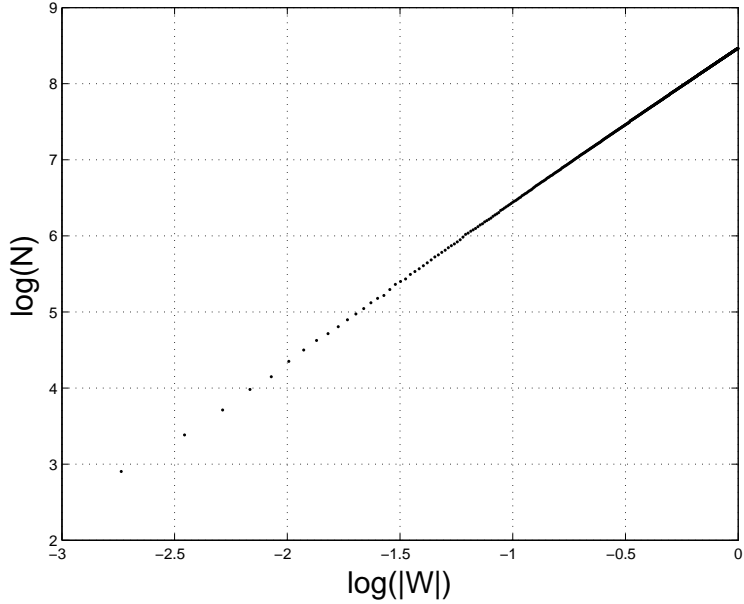}
\caption[fields]{\label{fig1} Number of vacua $N$ with $|{\cal W}_0|<w$
as a function of $w$ (in logarithmic units), for flux/torsion parameters
in the range $-70,\dots ,70$.}
\end{center}
\end{figure*}
In Figure~\ref{fig1} we have presented the result for $N(70,w)$, the
number of vacua in the flux/torsion range from $-70,\dots ,70$, as a
function of the superpotential value, $|W|$. The numerical value
of $x$ can be easily understood intuitively. Although we consider a
model with seven parameters, the requirement of small $\Re ({\cal W}_0)$
and small $\Im ({\cal W}_0)$ effectively fixes two of these parameters,
leading to a scaling law with power $5$. We also remark that the value
of $y$ close to $y=2$ corresponds to a nearly uniform random
distribution of ${\cal W}_0$ values in the complex ${\cal W}_0$
plane.\footnote{We thank Nuno Antunes for pointing this out to us.}

As can be seen from Figure~\ref{fig1}, $|{\cal W}_0|$ values of $0.01$
and smaller can be obtained. Our result gives an indication of what
fraction of vacua leads to a gauge coupling of $s\simeq 24$ as
suggested by gauge unification in the MSSM. Assuming the $E_8$
beta-function coefficient~\eqref{c} such a value for $s$ translates
into $x\simeq 5$ which, from Eq.~\eqref{xsol} (setting $k=1$ for
simplicity) implies $|{\cal W}_0|\simeq 1/14$ (or, equivalently, ${\rm
  log} |{\cal W}_0| \simeq -1.15$). This means
approximately a fraction of $10^{-3}$ of all vacua with $|{\cal
  W}_0|<1$ lead a gauge coupling sufficiently weak to be compatible
with gauge unification. For a condensing gauge group smaller than
$E_8$ or a value of $k$ smaller than one this fraction will decrease
accordingly.

We have also checked the case $r\neq 0$ without the
constraint~\eqref{simplecons} and the results are similar to the one
obtained for $r=0$.

Finally we would like to show a numerical proof of the validity of our
procedure (i.e. the use of the global SUSY condition 
${\cal W}_i(T_0,Z_0)={\cal W}_a(T_0,Z_0)=0$ with $|{\cal W}_0/M_0| \ll
1$ in order to find an approximate solution for the $T$ and $Z$ fields).
We have chosen, within this $r=0$ case, values for the flux/torsion 
parameters as follows
\begin{equation}
\label{param}
e=-7 \;,\  \epsilon=-4 \;, \ \mu=2 \;, \ \rho=5 \;, \ p=1 \;, \ q=2 \;,
\ \xi=-13 \;,
\end{equation}
for which Eqs~(\ref{r0}) give the field values
\begin{eqnarray}
z_0 & = & 2.598 \;, \nn \\
t_0 & = & 2.165 \;, \nn \\
\zeta_0 & = & 0.500 \;, \label{glob} \\
\tau_0 & = & 2.250 \;, \nn
\end{eqnarray} 
with $|{\cal W}_0|=0.167$. As it is illustrated in Figure~3, these
values are very close to the actual solution to the F-equations, given
by
\begin{eqnarray}
z_0 & = & 2.598 \;, \nn \\
t_0 & = & 2.164 \;, \nn \\
\zeta_0 & = & 0.520 \;, \label{loc}\\
\tau_0 & = & 2.300 \;. \nn
\end{eqnarray}  
Using the beta-function coefficient \eqref{c} and $k=2$,
the values for the dilaton field are found to be 
\begin{equation}
  \begin{aligned}
      s=& 23.190 \; ,\\ 
      \sigma=& 22.399 \; ,
  \end{aligned}
\end{equation}
which proves that phenomenologically viable values for the gauge
coupling can be obtained for reasonable values of the flux parameters.
\begin{figure*}[!htb]
\begin{center}
\includegraphics[width=10cm]{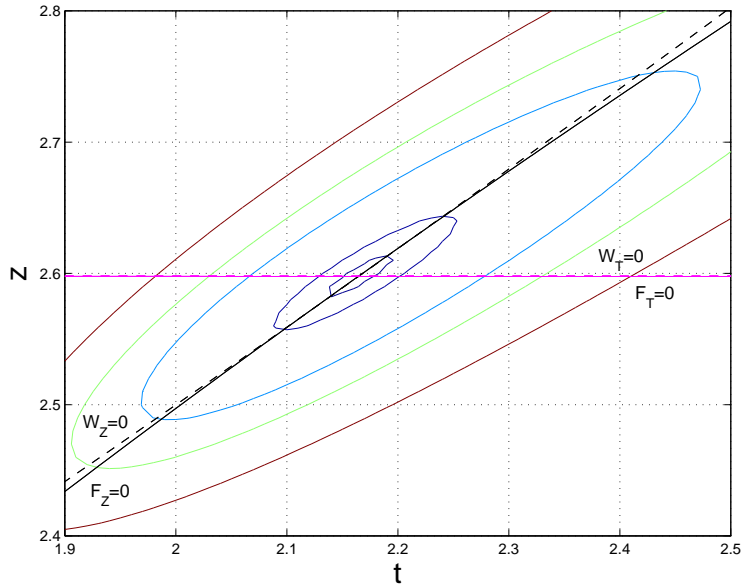}
\caption[fields]{\label{fig3}Contour plot of the potential, on the
  $(t,z)$ plane, for the example shown in the text, Eq.~\eqref{param}. The
  solid (dashed) lines are the $\Re (F_T)= \Re (F_Z)=0$ ($\Re (W_T)=
  \Re (W_Z)=0$) local (global) SUSY conditions. Note that the lines ${\rm Re}
  (W_T)=0$ and ${\rm Re} (F_T)=0$ coincide.}
\end{center}
\end{figure*}
We should add that we have computed the Hessian matrix for this
example and we have explicitly checked that all eigenvalues are
positive at the point where the F-equations vanish. This is quite
important as the potential at the minimum is negative, actually given
by $V=-3{\rm e}^K |{\cal W}_0|^2$, and very small, of the order of
10$^{-7}$. This means that, in a small region around the true minimum, the
potential will shift from negative to positive values and, this being
a multi-variable potential, it is easy to get mistaken about the real
position of the minimum.
For example, if we were to use the
global supersymmetric solution given by \eqref{glob}, due to its
closeness to the real one, \eqref{loc}, the no-scale cancellation mechanism would take place and
the scalar potential would read, at this point, $V=+3{\rm e}^K |{\cal
  W}_0|^2$. That is, we would predict a dS vacuum (of order 10$^{-7}$)
were, in reality, the only minimum in that region is AdS.

\vspace{.4cm}

Before we conclude this section we would like to make a few comments
about the general case, with an arbitrary numbers of moduli fields. As
it is evident from \eqref{suppot01} and \eqref{cpxflux}, the number of
parameters grows rapidly with the number of fields, making a numerical
search for vacua infeasible.  We have, therefore, not attempted to
extend the numerical analysis beyond the three-field STZ
model. However, the experience with this model shows that dropping the
constraint~\eqref{c2} is crucial in order to obtain small values of
$|{\cal W}_0|$ and we expect this to be true in general. Conversely,
multi-field models without the constraint~\eqref{c2} should be at
least as flexible as their three-field counterpart and should,
therefore, allow small $|{\cal W}_0|$ values without problems. From
our general argument relating globally and locally supersymmetric
vacua, this should then allow consistent, locally supersymmetric vacua
at weak coupling as determined by Eq.~\eqref{xsol}. As for the scaling
law~\eqref{Neq}, in the general case one would expect scaling powers
$x\simeq n-2$, where $n$ is the total number of parameters in the
model and $y\simeq 2$, leading to the same uniform random distribution in
the ${\cal W}$ plane which we have observed in the STZ model.

\section{Conclusion}
\label{sec:conclusions}

In this paper we have analyzed the vacuum properties of various
classes of heterotic models on certain manifolds with $\mbox{SU(3)}$
structure. After a review of the heterotic string on half-flat mirror
manifolds~\cite{GLM}, defined by~\eqref{Whf}, we
have derived the superpotential for a more general class of manifolds
with $\mbox{SU(3)}$ structure which were introduced in
Refs~\cite{DAFTV,AT,GLW}.  We have explicitly verified in these
models that the application of the heterotic Gukov-Vafa-Witten type
formula for the superpotential leads to the same result as an explicit
reduction of the ten-dimensional bosonic terms. The resulting
superpotential, which is given in Eqs~\eqref{suppot01}, \eqref{cpxflux},
resembles very much the one obtained in type IIB orientifold
compactifications suggesting that one may recover the flexibility of
type II models in the heterotic case.  These flux/torsion
superpotentials depend on $h^{(1,1)}$ K\"ahler moduli $T^i$ as well as
on $h^{(2,1)}$ complex structure moduli $Z^a$, but are independent of
the dilaton $S$. We have, therefore, supplemented our superpotential
with a contribution from hidden sector gaugino condensation in order
to stabilize the dilaton, which is again similar to the type IIB
constructions where non-perturbative terms need to be added in order
to fix the K\"ahler moduli.

\vspace{0.4cm}

We have first analyzed moduli stabilization for the models based on
half-flat mirror manifolds and have found a number of problems.
Generally in those models, $h^{(1,1)}+h^{(2,1)}-1$ axion directions
remain flat, and it is hard to achieve the large radius and large
complex structure limits as well as weak gauge coupling. In models
with additional magnetic torsion terms, the flat axion directions
are lifted and moderately weak coupling can be achieved, while
stabilizing field in the large radius and large complex structure
limits remains a problem. However, such models with additional 
magnetic torsion terms, although a plausible extension of half-flat
mirror models, cannot currently be derived within the heterotic string.

We have traced the root cause of the aforementioned problems to the
fact that the flux/torsion superpotential ${\cal W}$ is too simple to
allow for globally supersymmetric vacua. Consequently, we have
analyzed moduli stabilization for some generalised half-flat models
whose associated superpotential is significantly more complicated. We
have seen that consistent weak-coupling vacua can be obtained if the
flux/torsion superpotential has globally supersymmetric vacua with a
superpotential value ${\cal W}_0$ satisfying $|{\cal W}_0|\ll 1$.  The
value of the dilaton and, hence, the gauge coupling, is then directly
related to $|{\cal W}_0|$. We have verified that the superpotential for
the generalised half-flat models has indeed globally supersymmetric vacua
with all K\"ahler and complex structure moduli stable. However, the
requirement of small $|{\cal W}_0|$ turned out to be more subtle.  For
the standard embedding of the spin connection into the gauge group, the
resulting Bianchi identity $d{\hat H}=0$ for the NSNS form led to a
constraint~\eqref{c2} on the flux/torsion parameters which ruled out
the possibility of small ${\cal W}_0$, at least within the range of
flux/torsion parameters covered by our numerical scan.  However, for
more general gauge bundles, the constraint should be dropped and vacua
with small $|{\cal W}_0|$ can easily be obtained in this case. The number
of such vacua as a function of $|{\cal W}_0|$ for a simple model with
three fields, $S$, $T$ and $Z$, has been plotted in
Figure~\ref{fig1}. Using Eq.~\eqref{xsol} one can estimate that,
typically, the fraction of vacua that lead to a
sufficiently weak gauge coupling consistent with gauge unification,
is $10^{-3}$. Our results establish the existence
of consistent, weak-coupling AdS vacua within generalised heterotic
half-flat models.

\vspace{0.4cm}

In the light of these results, it is clearly desirable to get to a
better understanding of half-flat compactifications and their
generalizations, in particular with regard to the precise nature of
the manifolds involved, the rules for quantizing flux and torsion
parameters in those compactifications and the inclusion of gauge and
gauge matter fields. We will leave those tasks for future
publications.

\vspace{1cm}
\noindent
{\large\bf Acknowledgments} We would like to thank Nuno Antunes,
Kiwoon Choi, Jan Louis, Hans--Peter Nilles and Silvia Vaul\`a for
helpful discussions. BdC, AL and AM are supported by PPARC. SG is
supported by the JSPS under contract P03743.


\vskip 1cm
\appendix{\noindent\Large \bf Appendix}
\renewcommand{\theequation}{\Alph{section}.\arabic{equation}}
\setcounter{equation}{0}


\section{Ricci scalar for the ``extended half-flat'' manifolds}
\label{appA}

This section contains a generalization of the result obtained in the
appendices of Ref.~\cite{GLMW}, where the Ricci scalar for half-flat
manifolds mirror to Calabi-Yau with NS-NS fluxes was computed. Here we
will follow this calculation closely, by recalling the main identities
which remain valid, while pointing out the places where it differs
from the simple half-flat case.

To set the stage, let us briefly recall a few features of
manifolds with SU(3) structure. Such manifolds are characterized by
the existence of an almost complex structure with the associated
fundamental form $J$ and a $(3,0)$ complex form $\Ox$ which are
invariant under the action of the SU(3) structure group. More
concretely this means that the forms $J$ and $\Ox$ are covariantly
constant with respect to some connection $\nabla^{(T)}$, which in
general has a torsion. Decomposed into SU(3) representations
the torsion falls into five different classes $\cW_1 \; , \ \ldots ,
\cW_5$ which are given by
\begin{eqnarray}
  \label{torcls}
  d J & = & \cW_1 \Ox + \cW_4 \wg J+ \cW_3 \; ,\\
  d \Ox & = & \cW_1 J\wg J + \cW_5 \wg \Ox + \cW_2 \nn \; .
\end{eqnarray}
Since the torsion on manifolds with SU(3) structure measures the
departure from Calabi--Yau manifolds (which are Ricci flat) it is
clear that the Ricci scalar of the SU(3) structure manifolds depends
on their torsion. Thus, in order to compute the Ricci scalar, we will
need to know all the components of the torsion and for this we will
use equations \eqref{torcls} above and the relations
\begin{eqnarray}
  d\Ox & = & E_i\tox^i \; , \label{dO}\\
  dJ & = & (t^ip_{Ai})\bx^A-(t^iq_i^A)\ax_A \label{dJ} \; ,
\end{eqnarray} 
which are easily derived from \eqref{Omega} and
\eqref{extalg}, with the quantities $E_i$ defined in equation \eqref{Ei}. 
Note we have postulated that the basis forms in the above equations have
the same SU(3) properties as their Calabi--Yau counterparts. Hence,
$\ax_A$ and $\bx^A$ in Eq.~\eqref{extalg} are primitive and, consequently,
$\cW_4$ has to vanish. Moreover as $d \Ox$ in
Eq.~\eqref{dO} is a $(2,2)$ form, $\cW_5$ also vanishes. The other
torsion components $T_{1 + 2}$ and $T_3$ are found to be
\begin{eqnarray}
  (T_{1+2})_{\ax\bx\cx} & = & \frac{\bar E_i}{4 ||\Ox||^2} 
  (\tox^i)_{\ax \bx \ab \bb} \Ox^{\ab \bb}{}_{\cx} \; , \label{T12} \\
   (T_3)_{\ax\bx\cb} & = & -\frac{i}{2} t^i(p_{Ai} \bx^A_{\ax\bx\cb}
   - q_i^A \ax_{A\ax\bx\cb}) \label{T3} \; .
\end{eqnarray}
Here $||\Ox||^2 = \frac16 \Ox_{\ax \bx \cx} {\bar \Ox}^{\ax \bx \cx}$
is a function of the complex structure moduli which is related to the
K\"ahler potential for these fields via ${\cal V} ||\Ox||^2 =
e^{-K^{(Z)}}$, where ${\cal V}$ is the volume of the manifold. Note that, in
this case, the quantities $E_i$ are neither real nor constants as it
happened for the case of half-flat manifolds and thus one generically has
\begin{equation}
  T \in \cW_1 \oplus \cW_2 \oplus \cW_3 \; .
\end{equation}
However it is important to note that the nature of the indices of the
torsion components is the same as in the half-flat case, and the
torsion itself is still traceless. As a consequence, the expression of
the Ricci scalar in terms of the torsion obtained in \cite{GLMW} still
holds
\begin{equation}
  R = (T_{1+2})_{\ax \bx \cx} (T_{1+2})^{\ax \bx \cx} 
  - 6 (T_{1+2})_{\ax\bx\cx}(T_{1+2})^{\bx\cx\ax}
  + (T_3)_{\ax\bx\cb} (T_3)^{\ax\bx\cb} + {\rm c.c.}
\end{equation} 
To obtain the integrated Ricci scalar we perform the same steps as
in Ref.~\cite{GLMW} and, using the relations \eqref{ABC}, we
obtain 
\begin{eqnarray}
  \int_{\hat{X}}\sqrt{g}\, R & = & {\rm e}^{K^(Z)} (g^{ij} E_i\bar E_j 
  - 4 t^i t^j E_i\bar E_j) \nn \\ 
  && -\frac{1}{2} t^i t^j (p_{Ai} - q_i^C\cM_{CA}) 
  \big(\Im \cM^{-1} \big)^{AB} (p_{Bj}-q_j^D\bar\cM_{BD}) \; . 
\end{eqnarray} 
After taking into account various coefficients and rescalings, the
contribution of gravity to the potential in Einstein's frame can be
rewritten as
\begin{eqnarray}
  e^{-K}V_g & = & 8(g^{ij}E_i\bar E_j-4t^it^j E_i\bar E_j)\\
  && -4 e^{-K^{(Z)}} \big[\Im (\te_A)  - \Im (\tm^C) \cM_{CA}\big]\big(\Im
  \cM^{-1} \big)^{AB}\big[\Im (\te_B)  - \Im (\tm^D) \bar\cM_{DB}\big]\; . \nn
\end{eqnarray}

\section{Some useful results on special geometry}
\label{appB}

As it is well known, the moduli space of Calabi--Yau manifolds splits
into a product of two special K\"ahler manifolds, one for the
complexified K\"ahler class deformations and one for the complex
structure deformations. Since these geometries are at the heart of the
four-dimensional physics obtained from compactifications on
Calabi--Yau manifolds we review in this appendix some of the properties
of the special K\"ahler manifolds which we need in the main part of
the paper. We mainly follow Ref.~\cite{N2}.

The main feature of special K\"ahler manifolds is that their geometry
is completely determined in terms of a holomorphic function $\cH$,
called the pre-potential. In terms of the projective coordinates $\cX^P$,
where $P=0, \ldots , n$ ($n$ being the complex dimension of the
manifold), the pre-potential is a homogeneous function of degree two, that is,
it satisfies $\cX^P \cH_P = 2 \cH$, where $\cH_P =
\frac{\partial}{\partial \cX^P} \cH$. In fact, one does not always need
to rely on the pre-potential and it may be sufficient to work with
the period vector
\begin{equation}
  \label{Osec}
  \Oh = \left(
    \begin{array}[h]{c}
      \cX^P \\
      \cH_P \\
    \end{array}
    \right).
\end{equation}
Let us further introduce the symplectic inner product $<,>$ as
\begin{equation}
  \label{innp}
  <\Oh , \bar \Oh> \equiv \Oh^T \left(
  \begin{array}{cc}
    0 & - \mathbf{1} \\
    \mathbf{1} & 0 
  \end{array}
  \right) \bar \Oh = (\cH_P \bar \cX^P - \bar \cH_P \cX^P) \;.
\end{equation}
With this notation the K\"ahler potential can be written as
\begin{equation}
  \label{KsK}
  K = - \log{\big(i <\Oh, \bar \Oh> \big)} = - \log{\big[ i(\cH_P \bar
  \cX^P - \bar \cH_P \cX^P) \big]} \; ,
\end{equation}
while the K\"ahler metric is given by the usual formula
\begin{equation}
  g_{p \bar q} = \partial_p {\bar \partial}_{\bar q} K \; .
\end{equation}
Here the derivatives are with respect to the affine coordinates
 $X^p=\cX^p/\cX^0$, where $p,q,\dots = 1,\dots ,n$. It is also useful
 to introduce the K\"ahler covariant derivative of the periods $\Oh$
\begin{equation}
  \label{Up}
  e^{- K/2} U_q \equiv e^{- K/2} \left(
    \begin{array}{c}
      f_q^P \\
      h_{P \; q}\\
    \end{array}
    \right) = \nabla_p \Oh = (\partial_p + \partial_p K) \Oh \; .
\end{equation}
The period matrix $\cQ$, which is a complex symmetric matrix is now
required to satisfy 
\begin{equation}
  \label{pM}
  \begin{aligned}
    \bar \cH_P =& {\bar \cQ}_{PQ} {\bar \cX}^Q \; ,\\
    h_{P \; q} =& {\bar \cQ}_{PS} f_q^S \; .
  \end{aligned}
\end{equation}
It can be shown, see Ref~\cite{CDF}, that, in terms of the pre-potential $\cH$,
the period matrix has the form 
\begin{equation}
  \label{Nsg}
  \cQ_{PQ} = \bar \cH_{PQ} + 2i \frac{(\Im \cH)_{PR} (\Im \cH)_{QS}
  \cX^R \cX^S}{(\Im \cH)_{RS} \cX^R \cX^S} \; , 
\end{equation}
where we have denoted $\cH_{RS} = \frac{\partial}{\partial \cX^R}
\frac{\partial}{\partial \cX^S} \cH$.

With this one can prove the following relations
\begin{eqnarray}
  \label{N2rel}
  <\Oh , U_p>  & = & <\Oh , \bar U_{\bar p}> = 0  \; ,\nn \\
  g_{p \bar q} & = & -i <U_p, \bar U_{\bar q}> = -2 f_p^P \Im
  \cQ_{PQ} {\bar f}_{\bar q}^Q \; , \\
  f_p^P {\bar f}_{\bar q}^Q g^{p \bar q} & = & - \frac12 (\Im \cQ)^{-1 \;
  PQ} - e^K {\bar \cX}^P \cX^Q \; . \nn
\end{eqnarray}
In order to make this discussion less abstract let us first apply the
above formalism to the complex structure moduli space of Calabi--Yau
manifolds. The periods $\Oh$ are now determined by the holomorphic
$(3,0)$ form $\Ox$. Moreover the inner product \eqref{innp} becomes
now the inner product for three-forms on the Calabi--Yau
manifold. With this one immediately finds that the K\"ahler potential
\eqref{KsK} precisely reproduces the one from equation \eqref{kt}.
Finally, the K\"ahler covariant derivatives in equation \eqref{Up}
give the components of the $(2,1)$ forms in the basis
\eqref{norm3}. With these identifications it is easy to see that most
of the relations in equation \eqref{N2rel} are straightforward, the
only non-trivial ones involving the period matrix. Denoting the period
matrix by $\cM$ in this case and the indices $P, Q, \ldots$ by $A, B,
\ldots = 1 , \ldots , h^{(2,1)}$, the last relations in equation
\eqref{N2rel} become
\begin{eqnarray}
  \label{N2relcs}
  g_{a \bar b} & = & -2 f_a^A \Im \cM_{AB} {\bar f}_{\bar b} ^B \; , \nn \\
  e^{K} D_a Z^A D_{\bar b} {\bar Z}^B = f_a^A {\bar f}_{\bar b}^B g^{a \bar
  b} & = & - \frac12 (\Im \cM ) ^{-1 \; AB} - e^{K} {\bar Z}^A Z^B \; .
\end{eqnarray}

Finally we note that in the basis \eqref{norm3} the period matrix
$\cM$ can be found to be
\begin{eqnarray}
  \label{ABC}
  \int_{Y_3} \ax_A \wg * \bx^B & = & - (\Re \cM)_{AC} (\Im \cM)^{-1 \;
  CB} \; , \nn \\ 
  \int_{Y_3} \ax_A \wg * \ax_A & = & - (\Im \cM)_{AB} - (\Re \cM)_{AC}
  (\Im \cM)^{-1 \; CD} (\Re \cM)_{DB} \; , \\
  \int_{Y_3} \bx^A \wg * \bx^B & = & - (\Im \cM)^{-1 \; AB} \; . \nn
\end{eqnarray}

\vspace{.4cm}

While the pre-potential is typically a complicated
function it simplifies considerably in certain limits in moduli space,
such as large radius and large complex structure limits for the
K\"ahler and complex structure moduli spaces, respectively.
In those limits, the pre-potential is given by a cubic function
\begin{equation}
  \label{cubicF}
  \cH = -\frac{1}{6} \frac{d_{pqr} \cX^p \cX^p \cX^r}{\cX^0} \; ,
\end{equation}
Such a cubic pre-potential defines what is known as \emph{very special geometry}.
Writing the affine coordinates as
\begin{equation}
 X^p=\xi^p + i x^p\; ,
\end{equation}
one finds for the K\"ahler potential
\begin{equation}
  K=-\ln \left(\frac43 \cK \right) \; ,
\end{equation}
where
\begin{equation}
 \cK = \frac{i}{8}d_{pqr}(X^p-\bar{X}^p)(X^q-\bar{X}^q)(X^r-\bar{X}^r)
     = d_{pqr} x^i x^j x^k\; .
\end{equation}

\vspace{0.4cm}

From equation \eqref{Nsg} one can also define a period matrix in this
case and then the relations \eqref{N2rel} follow by straightforward
algebraic manipulations. There are a number of further very special
geometry relations which are useful in the main part of this
paper. First, let us define
\begin{equation}
  \label{Kidef}
  \cK_p=d_{pqr}x^q x^r\; , \qquad \cK_{pq}=d_{pqr} x^r\; .
\end{equation}
With this, the first derivatives of the K\"ahler potential $K$ with
respect to $X^p$, denoted by $K_p$ and the K\"ahler metric $K_{p \bar q}$
can be written as
\begin{eqnarray}
  K_p &=& -\frac{3i \cK_p}{2\cK} \; , \\
  g_{pq} & = & K_{p \bar q} = -\frac{3}{2} \left(\frac{\cK_{pq}}{\cK} 
    - \frac{3\cK_p \cK_q}{2 \cK^2}\right) \; . 
\end{eqnarray}
Defining fields $x_p=g_{p q} x^q$ with lowered indices it is easy
to show that 
\begin{equation}
 x_p = \frac{3\cK_p}{4\cK}\; ,\qquad x_p x^p = \frac{3}{4}\; .
\end{equation}
These formulae lead immediately to the ``no-scale'' relation
\begin{equation}
  \label{noscale}
  K^{p \bar q} K_p {\bar K}_{\bar q} = 3 \; .
\end{equation}

\vspace{0.4cm}

Finally we note that a typical flux superpotential, for example as it
arises from the Gukov-Vafa-Witten formula, can be written as
\begin{equation}
  W = e_P \cX^P-m^P\cH_P\; .
\end{equation}
Here $e_P$ and $m^P$ depend on the fluxes and can be either real
constants or can also depend holomorphically on other (super)fields in
the theory, but not on $X^p$, as we have seen in section \ref{extmod}.
For the cubic pre-potential~\eqref{cubicF} the dependence on the
physical degrees of freedom $X^p$ can be made explicit after setting
$\cX^0$ to one and we find
\begin{equation}
 W = e_0+e_p X^p + \frac{1}{2} d_{pqr} m^p X^q X^r 
 - \frac{m^0}{6} d_{pqr} X^p X^q X^r \; .
\end{equation}
Note that after transforming to the ``phenomenological'' convention
for $X^p$ by $X^p\rightarrow -iX^p$ (and after dropping an overall
factor of $-i$ from $W$ which is irrelevant) the constant and
quadratic terms in the above superpotential pick up a factor of $i$.

\section{Supergravity conventions in $d=4$ and stability of
  supersymmetric vacua}  
\label{appC}

In this appendix we summarize conventions and relevant formulae for
four-dimensional $N=1$ supergravity~\cite{WB}. Further, we present an
elementary proof that solutions to the F-equations are always stable
vacua.

The bosonic terms in the action of four-dimensional $N=1$ supergravity
coupled to chiral fields $(\phi^X)=(S,T^i,Z^a)$ read
\begin{equation}
  \label{4daction}
  S = -\frac{1}{\kappa_4^2}\int_{M_4} \sqrt{-g} \; d^4 x \; \left[
      \frac{1}{2} R + K_{XY} \partial_\mu \phi^X\partial^\mu
      \bar{\phi}^Y + V \right ] \; , 
\end{equation}
where $\kappa_4$ is the four-dimensional Newton constant. As usual, 
$K_{XY}=\partial_X \partial_Y K$ is the K\"ahler metric while the potential
$V$ is given by the standard formula
\begin{equation}
  \label{N1pot}
  V= e^K\left( K^{XY} F_X\bar{F}_Y - 3 |W|^2\right) \; ,
\end{equation}
where $K^{XY}$ is the inverse K\"ahler metric and the F-terms $F_X$
are defined by
\begin{equation}
 F_X=W_X + K_X W\; .
\end{equation}
Here a subscript $X$ denotes a derivative with respect to $\phi^X$, as
usual. Note that we have considered the chiral fields and the K\"ahler
potential to be dimensionless, while the superpotential has dimension one,
a convention which is convenient for the discussion of moduli fields and
in line with the formulae in the main part of the paper.

\vspace{0.4cm}

In this paper we were interested in supersymmetric vacua of the
potential~\eqref{N1pot}, that is, vacua which can be obtained by
solving the F-equations
\begin{equation}
 F_X=W_X+K_XW=0\; . \label{F}
\end{equation}
It is easy to show, from Eq.~\eqref{N1pot}, that solutions to the
F-equations indeed constitute extremal points of the potential $V$.
The cosmological constant, $V_0$, at such an extremal point is given by
\begin{equation}
 V_0=-3e^K|W|^2\; .\label{V0}
\end{equation}
Without fine-tuning (to make $W$ at the extremal point vanish) this
value will usually be negative and, hence, we are generically dealing
with AdS vacua. The stability of AdS vacua in gravity coupled to
scalar fields was analyzed a long time ago~\cite{BZ,AD} by
Breitenlohner and Freedman and, independently, by Abbott and Deser. They
found that such vacua are stable if all scalar field masses are larger
than a certain lower bound, which is basically given by the
cosmological constant $V_0$. Hence in AdS space, unlike in Minkowski
space, negative (square) masses do not necessarily indicate an
instability. In fact, it can be shown in general, \cite{DNP}, that this
bound is always satisfied for supersymmetric vacua of supergravity
theories and, hence, such vacua are always stable. We will now present
an elementary proof of this statement.

\vspace{0.4cm}

Let us first formulate the Breitenlohner-Freedman bound for a theory
with canonically normalized real scalars $\chi^i$ and a potential
$V=V(\chi^i)$.  We assume the potential has a stationary point at
$\chi^i=\chi^i_0$ with negative cosmological constant
$V_0=V(\chi^i_0)<0$. Define the mass matrix as
\begin{equation}
  M_{ij}=\frac{\partial^2V}{\partial\chi^i\partial\chi^j}(\chi^k_0)\; .
\end{equation}
According to Breitenlohner and Freedman this stationary point leads to a
stable, AdS vacuum if $a_i\geq 3V_0/2$, where $a_i$ are the eigenvalues
of $M$. A sufficient criterion for this to be the case is that
\begin{equation}
 \xi^TM\xi-\frac{3}{2}V_0\xi^T\xi\geq 0 \;, \label{crit} 
\end{equation}
for all vectors $\xi$ in field space. To see that this inequality
implies the one for the eigenvalues, choose $\xi$ to be the eigenvectors
of the mass matrix $M$.  It is useful for the application to
supergravity to re-write this criterion for non-canonical
kinetic terms $G_{ij}\partial_\mu\chi^i\partial^\mu\chi^j$, where $G_{ij}$ is
the metric on field space. Then, Eq.~\eqref{crit} takes the form
\begin{equation}
 \xi^TM\xi-\frac{3}{2}V_0\xi^TG\xi\geq 0\; ,\label{crit1}
\end{equation}
where the mass matrix $M$ is now, of course, defined with respect to
the non-canonical fields.

\vspace{0.4cm}

We would like to apply the criterion~\eqref{crit1} to the case of
four-dimensional $N=1$ supergravity with complex scalars $\phi^X$,
K\"ahler potential $K$ and superpotential $W$. We consider a solution
$\phi_0^X$ of the F-equations~\eqref{F}. This solution is
automatically a stationary point of the potential $V$ and it preserves
supersymmetry. The cosmological constant $V_0$ at
such a vacuum is given by Eq.~\eqref{V0}. From Eq.~\eqref{N1pot}
one finds for the second derivatives of $V$ at $F_X=0$ after a bit of
computation
\begin{eqnarray}
 \left. V_{XY}\right|_{F=0} &=& -e^K\bar{W}F_{XY} \;, \\
 \left. V_{X\bar{Y}}\right|_{F=0} &=& e^K\left[K^{Z\bar{T}}F_{ZX}\bar{F}_{\bar{T}\bar{Y}}
    -2K_{X\bar{Y}}|W|^2\right]\; ,
\end{eqnarray}
where $F_{XY}$ is the derivative of $F_X$ with respect to $\phi^Y$.
Combining these results (and taking care to convert real into complex
expressions) it is then straightforward to compute the LHS of the
criterion~\eqref{crit1} which takes the form
\begin{eqnarray}
 && 2V_{X\bar{Y}}\xi^X\xi^{\bar{Y}}+V_{XY}\xi^X\xi^Y+V_{\bar{X}\bar{Y}}\xi^{\bar{X}}
    \xi^{\bar{Y}}-\frac{3}{2}V_0K_{X\bar{Y}}\xi^X\xi^{\bar{Y}}\nonumber\\
 &=&e^K\left[2K^{Z\bar{T}}F_{ZX}\bar{F}_{\bar{T}\bar{Y}}\xi^X\xi^{\bar{Y}}-\bar{W}F_{XY}\xi^X\xi^Y
    -W\bar{F}_{\bar{X}\bar{Y}}\xi^{\bar{X}}\xi^{\bar{Y}}+\frac{1}{2}|W|^2K_{X\bar{Y}}
     \xi^X\xi^{\bar{Y}}\right]\nonumber\\
 &=&e^K\left(\frac{1}{\sqrt{2}}WK_{Z\bar{X}}\xi^{\bar{X}}-\sqrt{2}F_{ZX}\xi^X\right)
       K^{Z\bar{T}}\left(\frac{1}{\sqrt{2}}\bar{W}K_{\bar{T}Y}\xi^Y-\sqrt{2}\bar{F}_{\bar{T}
       \bar{Y}}\xi^{\bar{Y}}\right)\; .
\end{eqnarray}
The last line is obviously positive and, hence, the criterion is
satisfied. The conclusion is that any supersymmetric AdS vacuum of
four-dimensional $N=1$ supersymmetry satisfies the
Breitenlohner/Freedman criterion and, therefore, constitutes a stable
AdS vacuum.

\end{document}